\documentclass[final,5p,times,twocolumn]{elsarticle}

\usepackage{calrsfs,times}
\usepackage{graphicx,psfrag,enumerate}
\usepackage{amssymb}
\usepackage{amsmath}
\usepackage{calrsfs}
\usepackage{color, epsfig, epstopdf, wrapfig}
\usepackage{paralist}
\usepackage{float,cuted}
\usepackage{natbib}
\newtheorem{theorem}{Theorem}
\newtheorem{remark}{Remark}

\newtheorem{lemma}{Lemma}

\newtheorem{problem}{Problem}

\usepackage{amsmath}
\usepackage{accents}
\newlength{\dhatheight}
\newcommand{\doublehat}[1]{%
    \settoheight{\dhatheight}{\ensuremath{\hat{#1}}}%
    \addtolength{\dhatheight}{-0.35ex}%
    \hat{\vphantom{\rule{1pt}{\dhatheight}}%
    \smash{\hat{#1}}}}

\usepackage{color}
\usepackage[normalem]{ulem}
\usepackage{soul} 
\soulregister\cite7
\soulregister\ref7
\soulregister\pageref7

\journal{Automatica}
 \allowdisplaybreaks
\begin{document}

\begin{frontmatter}

\title{Detection and Mitigation of Biasing Attacks on Distributed Estimation Networks\tnoteref{mytitlenote}}
\tnotetext[mytitlenote]{This work was supported by the Australian Research Council and the University of New South Wales.}

\author[label1]{Mohammad Deghat}
\address[label1]{Department of Electrical and Electronic Engineering, The University of Melbourne, Parkville, VIC 3010, Australia}
\ead{m.deghat@unimelb.edu.au}

\author[label2]{Valery Ugrinovskii}
\ead{v.ougrinovski@adfa.edu.au}

\author[label1]{Iman Shames}
\address[label2]{School of Engineering and Information Technology, University of New South Wales at the Australian Defence Force Academy, Canberra, ACT 2600, Australia}
\ead{iman.shames@unimelb.edu.au}

\author[label3]{C\'{e}dric Langbort}
\address[label3]{Department of Aerospace Engineering and Coordinated Science Laboratory, University of Illinois at Urbana-Champaign, Urbana, IL 61801 USA}
\ead{langbort@illinois.edu}

\begin{abstract}
The paper considers a problem of detecting and mitigating biasing attacks on 
networks of state observers targeting cooperative state estimation
algorithms. The problem is cast within the recently developed framework of
distributed estimation utilizing the vector dissipativity approach. The
paper shows that a network of distributed observers can be endowed with an
additional attack detection layer capable of detecting biasing attacks and
correcting their effect on estimates produced by the network. An example 
is provided to illustrate the performance of the proposed distributed
attack detector.
\end{abstract}


\end{frontmatter}

\section{Introduction}
Recent developments in the area of networked control and
estimation have been increasingly focused on resilience of networked
control systems to intentional malicious input attacks aiming to compromise
stability and performance of control systems. Owing to the networked nature
of such control systems, typically not all measurements are available at
nodes of the network to allow efficient attack and fault
detection~\cite{Ferrari-2012, PDB-2015}; this has made    
distributed approaches particularly attractive. For instance,
\cite{TSSJ-2014} considers
distributed fault detection for second order dynamics at each node. Each
node has the model of the entire 
network or the model of its neighbourhood; in the latter case
interconnections of the neighbours to the agents outside the neighbourhood
are treated as undesirable disturbances to be rejected. A bank of fault
observers is constructed for each fault model. The situation is considered
where the network topology is uncertain, and can be captured as a
norm-bounded uncertain perturbation of the global network model.

Another fault detection algorithm is proposed in \cite{HWJZ-2011} which
considers a fault input to the plant with multiple randomly failing sensors
(random packet drop-out) for a discrete-time system. A discrete-time
system model is also considered in \cite{GHJ-2013}, and it is assumed that both
plant and sensors are subject to Markovian switching. The reference
considers a fault 
input to the plant and uses sensor information fusion from several nodes to
generate the residuals for fault detection.

A considerable progress on the problem has been achieved in~\cite{PDB-2013}
which not only considered the problem of residual generation for linear
systems given in a quite general descriptor form, but also has
characterized system vulnerabilities from the system theoretic perspective 
of attack input detectability. We also refer to~\cite{PDB-2015} where
connections have been drawn between the network topology and attack input
detectability.          

This paper considers the problem of detecting attacks on
consensus-based distributed estimation networks. The topic of distributed
estimation has gained considerable attention in the literature, in a bid to
reduce communication bottlenecks and improve reliability and fidelity of
centralized state observers. Filter cooperation and consensus ideas have
proved to be instrumental in the design of distributed state
observers~\cite{Olfati-Saber-2007,SWH-2010,U6}. At the same time,
consensus-based systems are particularly vulnerable to intentional attacks
since the compromised agents can interfere with the functions of the entire
network in a significant way~\cite{PBB-2012}. Uncertainty and noise
represent another challenge from the attack detection viewpoint --- state
observers are typically required in applications where uncertainty and
noise make accessing the system state difficult; this may allow the
attackers to remain undetected by injecting signals compatible with the
noise statistics~\cite{PDB-2013}. This motivates an increased interest in detection of rogue behaviours of state observers~\cite{MS-2016}. 
  
In this paper, we are concerned with resilience properties of a general class of
distributed state estimation networks considered, for example, 
in~\cite{SWH-2010,U6,LaU1}. The attack model assumes that the state observers
at the compromised nodes are driven by certain attack/fault
inputs. Referring to conventional false-data injection  
models~\cite{TSSJ-2015}, the model considered here is quite 
general, with several noteworthy features. Firstly, we consider the attacks that
force a rogue behaviour at the affected node by interfering with the
data processing algorithm. 
Similar to bias injection attacks considered in~\cite{TSSJ-2015}, the attack
inputs are not assumed to be constant and can include an uncertain
transient component to reflect the adversary's desire to make the attack
stealthy. 
The purpose of the attack under consideration is
to force the compromised node to produce biased state estimates and then
exploit interconnections within the network to propagate those
estimates across the network. 

In terms of resources required to launch an attack, the adversary does not
require any 
knowledge of the system model to carry out the attack, it relies on the
misappropriated observer to produce rogue messages which 
are then injected into other nodes of the network through existing network
communications. Our interest in this paper is in detecting and tracking
attack/fault inputs causing such a rogue behaviour. For this, the paper
proposes to introduce additional filters in the attack detectors at
every network node. The idea behind the introduction of such filters is to
differentiate imminently dangerous components of the attack
inputs from components whose effect is akin to that of `ever-present
disturbances'. Since distributed $H_\infty$ observers 
such as those considered in~\cite{SWH-2010,U6,LaU1} are constructed  to
attenuate disturbances, by design they are less
sensitive to exogenous `disturbance-like' inputs. On the other hand, 
low-frequency biasing inputs such as constant inputs do not directly fit within
the class of `finite-energy disturbances' typically considered within
the $H_\infty$ design. Such inputs may cause node observers to
produce biased state estimates, therefore they are of primary concern.
The proposed filters are to assist with monitoring the node observers to
detect biases caused by such inputs.

Tracking disturbances in a large-scale system is not uncommon. E.g.,
in~\cite{SA-2015} distributed integral action controllers were used for
averaging constant disturbances to enable all agents in the system to
synchronize to a common reference system governed by the averaged constant
disturbance. In contrast, in this paper we aim at tracking and suppressing
individual attack inputs applied at certain observer nodes, rather than
tracking an averaged attack vector. Additional filter dynamics introduced
in the fault detectors are 
instrumental to solve that task. 

From the viewpoint of fault detection/input estimation, the system subject
to attack is distributed itself. This problem setting is similar to
\cite{TSSJ-2014}, but is different from 
\cite{HWJZ-2011,GHJ-2013} which were focused on detecting faults in
the plant. Overall, the techniques developed in this paper have a
number of distinct features:
\begin{enumerate}[(a)]
\item
We are concerned with resilience of general multidimensional consensus-based
observers to biasing attacks that target estimation algorithms at
misappropriated observer nodes rather than sensor measurements or
communication channels. 
\item
The proposed attack detector is distributed and the node detectors
interact to detect and mitigate the attack. 
\item
Our approach utilizes methodologies of
vector dissipativity and $H_\infty$ estimation. This allows us to
accommodate 
uncertainties in the sensors and the plant
model.
\item
The proposed attack detectors utilize the same plant measurements and the same communication channels that are used by the networked observer to solve the original plant estimation task. But extra information needs to be communicated between the neighbour agents. This allows us to determine which 
of the node observers is compromised, without introducing additional
communication sensors or communication channels.
\end{enumerate}

The paper is an extended version of the conference
paper~\cite{DUSL1a}. Compared with the conference version, the paper has
been substantially extended. The present version includes complete proofs
and an example which were not included in the preliminary version. The
presentation has been substantially revised, to show that the results hold
under somewhat less restrictive design conditions. Also, two
new sections have been added which discuss the detectability conditions of the network and show that the proposed attack detectors
can be used for countering biasing attacks on a network of distributed
observers under consideration, providing the distributed estimation
algorithms under consideration with an additional level of resilience.   

The paper is organised as follows. In
Section~\ref{sec:distributed_estimation}, a background on distributed
consensus based estimation is presented, and the model of attack is
introduced. The class of biasing attacks is formally defined in
Section~\ref{Problem.formulation}, and
the attack detection problem is also formulated in that section. The main
results 
are given in Section~\ref{sec:lmi_design}, where sufficient conditions in
terms of coupled linear matrix inequalities are expressed to enable the
design of a networked attack detector. In Section~\ref{sect.resilience}, we
show that the outputs of the proposed attack detectors can be used for
correcting biased state estimates. This allows the system to remain
operational under attack, meeting the objective of resilient system design. 
Conditions on detectability of the network
are discussed in Section~\ref{sect.detectability}.
An illustrative example is presented in
Section~\ref{sec:simulations}, and finally some concluding remarks are given
in Section~\ref{sec:conclusion}.

\emph{Notation}: $\mathbf{R}^n$ denotes the real Euclidean $n$-dimensional vector space, with the norm  $\|x\|=(x'x)^{1/2}$; here the symbol $'$ denotes the transpose of a matrix or a vector.
The symbol $I$ denotes the identity matrix, and $0_{m\times n}$ denotes the
zero matrix of size $m\times n$. We will occasionally use $I$ and $0$ for notational convenience if no confusion is expected. 
The symbols $|\cdot|$ and $\mathrm{Re}(\cdot)$ denote respectively the magnitude and the real part of a complex number.
For real symmetric $n\times n$ matrices $X$ and $Y$, $Y>X$ (respectively, $Y\geq X$) means the
matrix $Y-X$ is positive definite (respectively, positive
semidefinite). 
$\mathrm{Ker}$ and $\mathrm{rank}$ denote the null-space and
rank of a matrix.
The notation $L_2[0, \infty)$ refers to the Lebesgue space of
$\mathbf{R}^n$-valued vector-functions $z(.)$, defined on the time interval
$[0, \infty)$, with the norm $\|z\|_2\triangleq\left(\int_0^\infty
  \|z(t)\|^2 dt \right)^{1/2}$ and the inner product $\int_0^\infty z_1'(t)
z_2(t) dt$. 
   
\section{Background: Continuous-time distributed estimation}
\label{sec:distributed_estimation}

Consider an observer network with $N$ nodes and a directed graph topology
$\mathbf{G} = (\mathbf{V},\mathbf{E})$ where 
$\mathbf{V}$ and $\mathbf{E}$ are the set of vertices and the set of edges (i.e, the subset of the set $\mathbf{V}\times \mathbf{V}$), respectively. 
Without loss of generality, we let $\mathbf{V}=\{1,2,\ldots,N\}$. 
The graph $\mathbf{G}$ is assumed to be directed, reflecting 
the fact that while node $i$ receives the information from node $j$, this
relation may not be reciprocal
. The notation $(j,i)$ denotes the edge
of $\mathbf{G}$  originating at node $j$ and ending at node $i$. It is assumed that
the nodes of the graph $\mathbf{G}$ have no self-loops, i.e.,
$(i,i)\not\in \mathbf{E}$. 
   
For each $i\in \mathbf{V}$, let $\mathbf{V}_i=\{j:(j,i)\in \mathbf{E}\}$ be the set of nodes supplying information to node
$i$. 
The cardinality of
$\mathbf{V}_i$, known as the in-degree of node $i$, is denoted $p_i$; i.e.,
$p_i$ is equal to the number of incoming edges for node $i$. Also,  $q_i$
will denote the number of outgoing 
edges for node $i$, known as the out-degree of node $i$.
Let $\mathbf{A}=[\mathbf{a}_{ij}]$ be the adjacency matrix of the
digraph $\mathbf{G}$, i.e., $\mathbf{a}_{ij}=1$ if $(j,i)\in \mathbf{E}$,
otherwise $\mathbf{a}_{ij}=0$.
Then, $p_i=\sum_{j=1}^N\mathbf{a}_{ij}=\sum_{j\in \mathbf{V}_i} \mathbf{a}_{ij}$, $q_i=\sum_{j=1}^N\mathbf{a}_{ji}$. 

A typical distributed estimation problem considers a plant described  by
the equation 
\begin{equation}
  \label{eq:plant}
  \dot x=Ax+B\xi(t), \quad x(0)=x_0, \quad x\in\mathbf{R}^n,
\end{equation}
governed by a disturbance input $\xi\in \mathbf{R}^m$. A network of
filters connected according to the graph $\mathbf{G}$ takes measurements of
the plant with the purpose to produce an estimate of $x$. It is assumed
that each filter takes measurements   
\begin{equation}\label{U6.yi}
y_i=C_ix+D_i\xi+{\bar D_i}\xi_i, 
\end{equation}
where $\xi_i(t)\in\mathbf{R}^{m_i}$ represents the measurement disturbance
at the local sensing node $i$, and processes them locally using an
information communicated by its neighbours $j$,
$j\in\mathbf{V}_i$. An underlying feature of the problem is that in general,
the pairs $(A,C_i)$ are not required to be detectable. This has an
implication that the nodes with undetectable pairs $(A,C_i)$ can only
obtain biased estimates of the plant, making cooperation between the
nodes a necessity. Requirements on local sensors and the network to enable
unbiased cooperative networked state estimation have been considered in the 
recent literature; e.g., see~\cite{Doostmohammadian2013,U7b-journal,BC-2014}. 

Depending on the nature of the disturbances $\xi$,
$\xi_i$, cooperative processing of measurements can be done using
Kalman~\cite{Olfati-Saber-2007}, $H_\infty$~\cite{SWH-2010,U6,LaU1}
filters, etc. Many of the existing algorithms utilize networks of cooperating
filters, each producing an estimate $\hat x_i$ of the state $x$ using an
observer of the form  
\begin{eqnarray}
    \dot{\hat x}_i=A\hat x_i + L_i(y_i(t)-C_i\hat x_i)+
    K_i\sum_{j\in \mathbf{V}_i}H(\hat x_j-\hat x_i), 
  \label{UP7.C.d.unbiased} \\
 \hat x_i(0)=0; \nonumber
\end{eqnarray}
here the matrices $L_i$, $K_i$ are the parameters of the filter. Each
filter combines processing of innovations obtained from local measurements
with feedback from its neighbours, captured in the last term
(\ref{UP7.C.d.unbiased}) where the neighbours' estimates are
$\hat x_j$, $j\in \mathbf{V}_i$. The matrix $H$
determines what information about $\hat x_j$ is shared between the nodes. 
For simplicity of presentation, we assume that communication channels
between the nodes are ideal, and node $i$ receives the precise value of
$H\hat x_j$, and that the matrices $H$ are identical across the network.   
More general formulations 
which allow for disturbances in communication channels and
heterogeneity in communicated information 
can be easily
accommodated within our approach, as they do not bring additional technical
challenges. 

The general problem of distributed estimation is to determine
estimator gains $L_i$ and $K_i$ in \eqref{UP7.C.d.unbiased} to ensure the
filter internal stability and acceptable filtering performance against
disturbances. Therefore, from the system resilience viewpoint it is of
interest to consider the situation where one or several nodes of the
network of observers \eqref{UP7.C.d.unbiased} are subject to an attack
whose aim is to interfere with these filtering performance objectives.  
A most common scenario of such an attack considered in the literature
involves the attacker 
tempering with the measurements and/or
communications between the nodes. 
In contrast, we consider the situation where the attacker mounts an
attack on the observer dynamics directly. That is, we consider the
scenario where some of the nodes are misappropriated by the attacker and,  
in lieu of (\ref{UP7.C.d.unbiased}), generate their estimates according to  
 \begin{eqnarray}  
    \dot{\hat x}_i&=&A\hat x_i + L_i(y_i(t)-C_i\hat x_i) \nonumber \\
&& +K_i\sum_{j\in
      \mathbf{V}_i}H(\hat x_j-\hat x_i)+F_if_i, \qquad \hat x_i(0)=0.
  \label{UP7.C.d}
\end{eqnarray}
Here $F_i\in\mathbf{R}^{n\times n_f}$ is a constant ``fault entry matrix'' (e.g., see \cite{Patton1997}) and $f_i\in \mathbf{R}^{n_f}$ is the unknown signal representing an attack input. The gains of the observers and the network topology are not affected by the attacker and are assumed to be fixed
(cf.~\cite{Doostmohammadian2013}).
From now on, our focus is exclusively on
the network of observers 
(\ref{UP7.C.d}), although our approach to detection and mitigation of
biasing attacks can be readily applied to other mentioned distributed state
estimation algorithms as well. For instance, if the communication link
between node $i$ and $k: k\in\mathbf{V}_i$ is under attack such that node
$k$ instead of receiving $\hat x_k$ receives a biased estimate of $\hat
x_k+\ell_{ik}$ where $\ell_{ik}$ is an attack signal, then this situation
is still captured by the biased estimator model~\eqref{UP7.C.d} in which
$F_i=K_iH$ and $f_i=\ell_{ik}$. Therefore, the analysis presented in the
paper is applicable to this type of attack as well.

The class of attacks considered in this paper does not contain attacks that
cause nodes or links to fall out of the network. Our consideration is that
the objectives of the biasing attacker are different from the objectives of
a jamming attacker. Attacking links to fail is a kind of DoS attack, and
these attacks disrupt the normal flow of information within the network. In
contrast, the biasing attacker who misappropriates a node benefits from
integrity of the network links, since it uses them to spread the biased
$\hat x_i$ across the network. Therefore the analysis in the paper is
carried out under the assumption that it is not in the 
attacker's interests to block network links. Attack stealthiness
considerations also support this assumption. While jammers act openly to
block communication links or sensing nodes\footnote{Generally, it is quite
  difficult for the jammer to remain stealthy~\cite{LaU3}.}, we
consider 
that the intention of a biasing attacker is to remain hidden, in order to
inject the false data for as long as possible. Unusual patterns in nodes
and links failures will likely to prompt maintenance which may reveal the
attacker. Thus it may be risky for the biasing attacker to disrupt
connectivity if it wishes to remain stealthy.~\label{VU.connectivity.remark}

\begin{figure}[t]
\psfrag{R}{$f_i$}
\psfrag{nu}{$-\nu_i$}
\psfrag{Y}{$\hat f_i$}
\psfrag{H}{$G_i(s)$}
\psfrag{G}{$\frac{1}{s}$}
\psfrag{+}{$+$}
\psfrag{-}{$-$}
  \centering
  \includegraphics[width=0.45\textwidth]{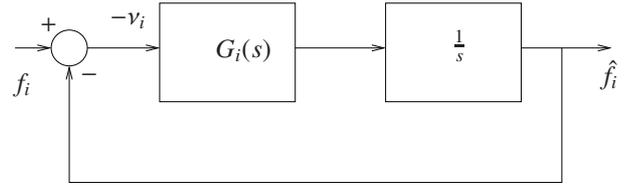}
  \caption{An auxiliary `input tracking' model.}
  \label{tracker}
\end{figure}

\section{Problem formulation}\label{Problem.formulation}

To be concrete, from now on we build the
presentation around the distributed $H_\infty$ cooperative estimation
problem~\cite{U6,LaU1}, although the approach to bias attack detection
proposed in this paper is general enough to allow extensions to other types
of filters in an obvious manner. In line with the disturbance model
considered in~\cite{U6,LaU1}, it will be assumed throughout the paper that the
disturbances $\xi$, $\xi_i$ belong to $L_2[0,\infty)$. This assumption
suffices to guarantee that equation (\ref{eq:plant}) has an
$L_2$-integrable solution on any finite time interval $[0,T]$, even when 
the matrix $A$ is unstable.  

\subsection{Admissible biasing attacks}\label{attack.model} 

We now present a class of biasing attacks on misappropriated nodes of the
filter (\ref{UP7.C.d}) that will be considered in this paper.  
First consider a class of attack input signals $f_i(t)$, $t\ge 0$, of the form
\begin{equation} 
  \label{decomp}
f_i(t)=f_{i1}(t)+f_{i2}(t),  
\end{equation}
where the Laplace transform of $f_{i1}(t)$, $f_{i1}(s)$, is such that
$f_{i1}^\infty\triangleq \sup_{\omega}\|\omega f_{i1}(j\omega)\|^2<\infty$ and
$f_{i2}\in L_2[0,\infty)$.
In particular, this class includes attack inputs whose Laplace transform is rational and has no more than one pole at the origin, with the remaining poles located in the open left half-plane of the complex plane. This class of inputs will be denoted $\mathcal{F}$. It
includes as a special case bias injection attack inputs
consisting of a steady-state component and an exponentially decaying
transient component generated by a low pass filter introduced
in~\cite{TSSJ-2015}.

It is easy to show that there
exists a proper square $n_f\times n_f$ transfer function\footnote{A proper
  transfer function 
  (respectively, a strictly proper transfer function) is a transfer function
  in which the degree of numerator does not exceed (respectively, is less
  than) the degree of the denominator.} $G_i(s)$ for which the system in 
Fig.~\ref{tracker} is stable and 
\begin{equation}\label{f-eta} 
 \int_0^\infty\|f_i-\hat f_i\|^2dt<\infty  
\end{equation}
for all $f_i$ with the properties stated above. Indeed,
we can select $G_i(s)=\frac{N(s)}{D(s)}I$, for which the system in
Fig.~\ref{tracker} is stable; here $N(s)$ and $D(s)$ are real polynomials with
$\mathrm{deg}(N)\le \mathrm{deg}(D)$. It follows from stability of the
system in Fig.~\ref{tracker} that $j\omega D(j\omega)+N(j\omega)\neq 0$
$\forall \omega$.  This conclusion trivially follows from the Nyquist
criterion. Hence, $a\triangleq \sup_\omega \left|\frac{j\omega
    D(j\omega)}{j\omega D(j\omega)+N(j\omega)}\right|^2
<\infty$. Then\footnote{Here,  
  $\|\cdot\|$ is the induced norm of a matrix.}  
\begin{eqnarray}
&& \frac{1}{2\pi}\int_{-\infty}^{+\infty}
\|(I+\frac{1}{j\omega }G_i(j\omega))^{-1}f_i(j\omega )\|^2d\omega \nonumber
\\  
&& \le 
f_{i1}^\infty
\int_{-\infty}^{\infty}\left|\frac{D(j\omega)}{j\omega
    D(j\omega)+N(j\omega)}\right|^2d\omega  \nonumber \\
 && + 
a\int_{-\infty}^{\infty}\|f_{i2}(j\omega)\|^2d\omega < \infty.
\label{eq:1}
\end{eqnarray}
Define $\nu_i=\hat f_i-f_i$. Denoting the Laplace transforms of $f_i$ and
$\nu_i$ as $f_i(s)$ and $\nu_i(s)$ respectively, and noting that
\[
\nu_i(s)=-(I+\frac{1}{s}G_i(s))^{-1}f_i(s),
\]
we conclude from (\ref{eq:1}) that (\ref{f-eta}) is satisfied. 

Furthermore, if $G_i(s)$ is selected so that
\begin{eqnarray}
\lim_{s\to 0}
\|(I+\frac{1}{s}G_i(s))^{-1}\|=0, \label{G} 
\end{eqnarray}
then for all inputs $f_i\in \mathcal{F}$,
\begin{equation}
  \label{eq:2}
  \lim_{t\to\infty}\|f_i(t)-\hat f_i(t)\|= 0.
\end{equation}
The proof of this fact is given in the Appendix. 

In summary, we have observed that a large class of biasing inputs (which
includes bias injection attack inputs introduced
in~\cite{TSSJ-2015}) can be represented as
\[
f_i=\hat f_i-\nu_i,
\]
where $\nu_i$ is an $L_2$ integrable discrepancy
between the attack input $f_i$ and its `model' $\hat f_i$. For this, the
model generating transfer function $G_i(s)$ needs to satisfy very mild
assumptions - it must be proper, and the closed loop system in
Fig.~\ref{tracker} must be stable. Other than that, $G_i(s)$ can be chosen
arbitrarily. This allows us to proceed assuming formally that a
collection of transfer functions $G_i(s)$, $i=1, \ldots, N$,  with the
above properties has been selected, and a set of biasing inputs $f_i$ is
associated with this selection of transfer functions consisting of all
signals $f_i$ for which (\ref{f-eta}) holds. We will refer to the inputs from
this set as  \emph{admissible} biasing inputs. Clearly, such set is
quite rich; as we have shown, it 
subsumes all inputs (\ref{decomp}) and, consequently, the input set
$\mathcal{F}$ and biasing attack inputs defined in~\cite{TSSJ-2015}. In
addition, $L_2$-integrable inputs $f_i$ which represent attack inputs with
a limited energy resource~\cite{TSSJ-2015} also belong to the set of
admissible inputs since they are trivially represented in the form
(\ref{decomp}). It must be stressed that even though $G_i(s)$ is selected,
the details of admissible  biasing inputs, e.g., the asymptotic steady-state
value or the shape of the transient, remain unknown to the designer.       

We conclude this section by presenting a state space form of the system in
Fig.~\ref{tracker} which will be used in the sequel. 
Since $G_i(s)$ is proper, the transfer function $\frac{1}{s}G_i(s)$ is
strictly proper. Hence, a state space realization for the system in
Fig.~\ref{tracker}, e.g., the
minimal state space realization, is of the form
\begin{eqnarray}
&&\dot\epsilon_i = \Omega_i\epsilon_i+\Gamma_i \nu_i, \qquad
\epsilon_i(0)=0, \label{Om.sys.general} \\ 
&&\hat f_i= \Upsilon_i\epsilon_i, \nonumber
\end{eqnarray}
where $\nu_i=\hat f_i-f_i$ is an $L_2$-integrable input. For example, for
the system $G_i(s)=\frac{d_i}{s+2\beta_i}I$, we can let $\epsilon_i\in
\mathbf{R}^{2n_f}$, and 
\begin{equation}\label{omega}
\Omega_i=\left[\begin{array}{cc} 0 & I \\ 0 & -2\beta_i
    I\end{array}\right], \quad \Gamma_i=\left[\begin{array}{c} 0  \\
          -d_iI\end{array}\right], \quad \Upsilon_i=[I~0].
\end{equation}
In what follows, the state space model (\ref{Om.sys.general}) will be used
in the derivation of attack detectors, and the sufficient conditions for
attack detection proposed in the paper will include the parameters of the
model (\ref{Om.sys.general}). Some trials may be required in order to
select these parameters to obtain satisfactorily performing detectors.

\subsection{The distributed attack detector}

The objective of the paper is to design a distributed attack detection
system which is capable of tracking and suppressing admissible attack
inputs
. To achieve this task, we first
summarize the information about the network available at each node, which
will be used by the attack detectors. This information consists of the pair
of innovation output signals 
\begin{eqnarray}
  \zeta_i&=&y_i-C_i\hat x_i =C_i(x-\hat x_i) + D_i\xi+{\bar D_i}\xi_i, 
                                              \label{out.y} \\
  \bar\zeta_i&=&\sum_{j\in \mathbf{V}_i} H(\hat x_j-\hat x_i).
                                              \label{out.c}
\end{eqnarray}
The idea behind introducing these outputs is as follows. If node $i$ is
under attack, then its predicted sensor measurement $C_i\hat x_i $ is
expected to be biased, compared to the actual measurement $y_i$. This must
lead to a significant difference between these two signals, i.e., we must
expect a large energy in $\zeta_i$. Likewise, the plant state estimate $\hat
x_i$ at the misappropriated node $i$, is expected to deviate, at
least during an initial stage of the attack, from the estimates produced at
the neighbouring nodes. Thus, the variable $\bar\zeta_i$ describing
dynamics of the disagreement between node $i$ and its neighbours is
expected to differ from similar variables produced by network nodes not
affected by the attack. This motivates using the innovation signals
(\ref{out.y}), (\ref{out.c}) as inputs to
the attack detector. They can be
readily generated at node $i$; computing them only requires the
local measurement $y_i$, the local estimate $H\hat x_i$ computed by the
observer at node $i$ and the neighbours' signals $H\hat x_j$,
$j\in\mathbf{V}_i$, available at node $i$; see (\ref{UP7.C.d}). 

Let $e_i=x-\hat x_i$ be the local estimation error 
at node $i$. Using (\ref{eq:plant}) and (\ref{UP7.C.d}), it is 
straightforward to verify that each error $e_i$ satisfies the
following equation: 
 \begin{eqnarray}
    \dot{e}_i&=&(A - L_iC_i)e_i+K_i\sum_{j\in
      \mathbf{V}_i}H(e_j-e_i) \nonumber \\ &&+(B-L_iD_i)\xi-L_i{\bar
      D_i}\xi_i-F_if_i, \quad  e_i(0)=x_0. \qquad \label{e} 
\end{eqnarray}
Combine the system (\ref{e}) with the auxiliary input tracking model 
(\ref{Om.sys.general}):  
\begin{eqnarray}
    \dot{e}_i&=&(A - L_iC_i)e_i+K_i\sum_{j\in
      \mathbf{V}_i}H(e_j-e_i) -F_i\Upsilon_i\epsilon_i \nonumber \\ 
&+&(B-L_iD_i)\xi-L_i{\bar
      D_i}\xi_i+F_i\nu_i, \quad  e_i(0)=x_0,  \nonumber \\
      \dot \epsilon_i&=&\Omega_i \epsilon_i+\Gamma_i \nu_i \quad
      \epsilon_i(0)=0. \label{e.ext.nu.om} 
\end{eqnarray}
Here, we have used the relation $f_i=\Upsilon_i\epsilon_i-\nu_i$;
see (\ref{Om.sys.general}). 
The resulting system (\ref{e.ext.nu.om}) equipped with the outputs
(\ref{out.y.1}), (\ref{out.c.1}) can be regarded as an uncertain system
governed by 
$L_2$-integrable inputs $\xi$, $\xi_i$ and $\nu_i$. Each such system is
interconnected with its neighbours via inputs $e_j$, and the collection of
all such systems represents a large-scale system. The innovations
(\ref{out.y}), (\ref{out.c}) can be regarded as outputs of this large-scale
system since they can be written in terms of the estimation errors as 
\begin{eqnarray}
  \zeta_i&=&C_ie_i + D_i\xi+{\bar D_i}\xi_i, 
                                              \label{out.y.1} \\
  \bar\zeta_i&=&-\sum_{j\in \mathbf{V}_i} H(e_j-e_i).
                                              \label{out.c.1}
\end{eqnarray}

We propose the following distributed $H_\infty$ 
observer for the large-scale system (\ref{e.ext.nu.om}) as an attack
detector for the observer network (\ref{UP7.C.d}). The detector
is to utilize the outputs (\ref{out.y}), (\ref{out.c}) of the system
(\ref{UP7.C.d}) (equivalently, the outputs  (\ref{out.y.1}), (\ref{out.c.1})) 
to estimate dynamics $e_i$ and $\epsilon_i$ of the system (\ref{e.ext.nu.om})
while attenuating the disturbances $\xi$, $\xi_i$ and $\nu_i$, 
$i=1,\ldots,N$. The proposed detector is therefore as follows: 
\begin{eqnarray}
    \dot{\hat{e}}_i&=&(A - L_iC_i)\hat{e}_i+K_i\sum_{j\in
      \mathbf{V}_i}H(\hat{e}_j-\hat{e}_i) - F_i\Upsilon_i\hat\epsilon_i \nonumber \\ 
    &+&\bar L_i(\zeta_i-C_i\hat{e}_i)+\bar K_i\left(\bar\zeta_i+\sum_{j\in
      \mathbf{V}_i}H(\hat{e}_j-\hat{e}_i)\right), \nonumber \\
  \dot{\hat\epsilon}_i &=& 
\Omega_i \hat\epsilon_i +
    \check L_i(\zeta_i-C_i\hat{e}_i)+\check K_i\left(\bar\zeta_i+\sum_{j\in 
      \mathbf{V}_i}H(\hat{e}_j-\hat{e}_i)\right), \nonumber \\
\hat{\varepsilon}_i&=& \Upsilon_i\hat{\epsilon}_i, \label{ext.obs.nu.1.om} \\
&& \hat{e}_i(0)=0, \quad \hat\epsilon_i(0)=0. \nonumber 
\end{eqnarray}
The coefficients $\bar L_i$, $\bar K_i$, $\check L_i $, $\check K_i$ are to be
found to ensure that the output $\hat{\varepsilon}_i=
\Upsilon_i\hat{\epsilon}_i$ of the system (\ref{ext.obs.nu.1.om})
tracks the output $\hat f_i$ of the auxiliary system
(\ref{Om.sys.general}). Since 
$\hat
f_i$ converges to $f_i$ in the $L_2[0,\infty)$ sense, we propose that $\hat
\varepsilon_i$ is to be used as a residual variable indicating
whether the attack is taking place.

\begin{remark}\label{comp.cost}
The proposed attack detector requires each node to dynamically update two
other vectors, namely $\hat{e}_i$ and $\hat{\epsilon}_i$. Thus, in all,
each node will require updating an augmented vector whose dimension is
$2n + n_f$. This potentially increases the computational burden on the
filtering nodes. This is the price of dynamically estimating the state observer
error $e_i$ and $\epsilon_i$. In a typical distributed state estimation
scenario, state estimation errors are not observed. However, in our problem
concerned with resilient estimation, we require additional variables to
detect and track changes in the observer dynamics and to mitigate
the effect of the attack.  
\end{remark}

To formalize the above idea, introduce the error vectors $z_i=e_i-\hat{e}_i$,
$\delta_i=\epsilon_i-\hat\epsilon_i$ for the attack detector system
(\ref{ext.obs.nu.1.om}). It can be seen from (\ref{e.ext.nu.om}) and
(\ref{ext.obs.nu.1.om}) that the evolution of these
error vectors is governed by the following equations 
 \begin{eqnarray}
    \dot z_i&=&(A - L_iC_i)z_i+K_i\sum_{j\in
      \mathbf{V}_i}H(z_j-z_i) - F_i\Upsilon_i\delta_i \nonumber \\ 
    &-&\bar L_iC_iz_i+\bar K_i\sum_{j\in
      \mathbf{V}_i}H(z_j-z_i) \nonumber \\ &+&(B-L_iD_i)\xi-L_i{\bar
      D_i}\xi_i+F_i\nu_i -\bar L_i D_i\xi-\bar L_i {\bar
      D_i}\xi_i,  \nonumber \\
    \dot{\delta}_i&=&\Omega_i\delta_i -
    \check L_iC_iz_i+\check K_i\sum_{j\in 
      \mathbf{V}_i}H(z_j-z_i) \nonumber \\
&-&\check L_iD_i\xi-\check L_i{\bar
      D_i}\xi_i +\Gamma_i \nu_i, \nonumber \\  
\varpi_i&=& \hat f_i-\hat\varepsilon_i= \Upsilon_i\delta_i,
\label{ext.error.0} \\
&&  z_i(0)=x_0, \quad \delta_i(0)=0. \nonumber 
\end{eqnarray}
Using the notation $\tilde L_i=L_i+\bar L_i$, $\tilde
K_i=K_i+\bar K_i$, (\ref{ext.error.0}) can be simplified as   
 \begin{eqnarray}
    \dot z_i&=&(A - \tilde L_iC_i)z_i+\tilde K_i\sum_{j\in
      \mathbf{V}_i}H(z_j-z_i) - F_i\Upsilon_i\delta_i \nonumber \\ 
&+&(B-\tilde
    L_iD_i)\xi-\tilde L_i{\bar
      D_i}\xi_i+F_i\nu_i, \quad  z_i(0)=x_0,  \nonumber \\
    \dot{\delta}_i&=&\Omega_i\delta_i -
    \check L_iC_iz_i+\check K_i\sum_{j\in 
      \mathbf{V}_i}H(z_j-z_i) \nonumber \\
&-&\check L_iD_i\xi-\check L_i{\bar
      D_i}\xi_i +\Gamma_i \nu_i
, \quad  \delta_i(0)=0.
\label{ext.error} 
\end{eqnarray}

Our design objective can formally be expressed as the 
problem
concerned with asymptotic behaviour of the system (\ref{ext.error}). 
   
\begin{problem}[The distributed $H_\infty$ detector design
  problem]\label{Prob1}   
The distributed attack detection problem
is to determine $\bar L_i$, $\bar K_i$, $\check L_i$, $\check K_i$ for the
distributed attack detector (\ref{ext.obs.nu.1.om}) which ensure 
that the following properties hold:  
\begin{enumerate}[(i)]
\item
The large-scale system (\ref{ext.error}) is
internally stable. That is, the disturbance and attack-free 
large-scale system
 \begin{eqnarray}
    \dot z_i&=&(A - \tilde L_iC_i)z_i+\tilde K_i\sum_{j\in
      \mathbf{V}_i}H(z_j-z_i) - F_i\Upsilon_i \delta_i ,  \nonumber \\
    \dot{\delta}_i&=&\Omega_i\delta_i -
    \check L_iC_iz_i+\check K_i\sum_{j\in 
      \mathbf{V}_i}H(z_j-z_i), \label{e.2} \\
	&& z_i(0)=x_0,  \quad  \delta_i(0)=0, \nonumber
\end{eqnarray}
must be asymptotically stable. 
\item 
In the presence of $L_2$-integrable disturbances and admissible biasing
inputs, the system (\ref{ext.error}) achieves a guaranteed 
level of $H_\infty$ disturbance attenuation:  
\begin{equation}\label{objective.i.1}
\sup_{x_0, \mathbf{w}\neq 0}\,
\frac{\int_0^\infty\sum_{i=1}^N(\delta_i'Q_i\delta_i+ z_i'\bar Q_i z_i) dt}
{\|x_0\|^2_P+\sum_{i=1}^N\|\mathbf{w}_i\|_2^2}
\le \gamma^2, 
\end{equation}
where $Q_i=Q_i'>0$, $\bar Q_i=\bar Q_i'\ge 0$ are given matrices,
$\|x_0\|^2_P=x_0'Px_0$, $P=P'>0$ is a fixed matrix to be determined later,
$\mathbf{w}_i\triangleq 
[\xi',\xi_i',\nu_i']'$, $\mathbf{w}\triangleq[\mathbf{w}_1',\ldots,
\mathbf{w}_N']'$,  and $\gamma>0$ 
is a constant. 
\end{enumerate}
\end{problem}

Properties (i) and (ii) reflect a desirable behaviour of the attack
detector. Indeed, it follows from (\ref{objective.i.1}) that each attack
detector output variable $\hat{\varepsilon}_i=\Upsilon_i\hat\epsilon_i$
provides an $H_\infty$ estimate of $\hat f_i$. We now show that for
admissible attacks, this output converges to $f_i$, and hence it can be
used as a residual output indicating whether an admissible attack is taking place. 

\begin{theorem}\label{L2.tracking}\label{assympt.tracking}
\begin{enumerate}[(i)]
\item
Suppose $f_i$ are admissible biasing inputs 
and the distributed
networked attack detector (\ref{ext.obs.nu.1.om}) 
is such that 
condition (\ref{objective.i.1}) holds.
Then $\int_0^\infty \|\hat\varepsilon_i- f_i\|^2dt<\infty$ $\forall i$.
\item
Furthermore, if in addition the disturbance and attack-free 
large-scale system (\ref{e.2}) is asymptotically stable, and also
(\ref{objective.i.1}) holds with $\bar Q_i>0$, then $\lim_{t\to\infty} z_i= 0$, 
$\lim_{t\to\infty} \|\hat \varepsilon_i-f_i\|= 0$ for all biasing inputs
$f_i\in\mathcal{F}$. 
\end{enumerate}
\end{theorem}


\emph{Proof: } To prove statement (i), let $\bar\sigma\triangleq
\max_{i}\|\Upsilon_i\|^2$, and 
$\sigma>0$ be a constant such that $Q_i>\sigma I$ $\forall i$. Then, since
for any admissible attack input $f_i$, $\nu_i=\hat f_i-f_i$ is
$L_2$-integrable (see~(\ref{f-eta})), we have 
\begin{eqnarray}
\lefteqn{\sum_{i=1}^N\int_0^\infty \|\hat\varepsilon_i- f_i\|^2dt} && \nonumber\\
&&\le 2\int_0^\infty \sum_{i=1}^N\|\varpi_i\|^2dt 
+ 2 \int_0^\infty \sum_{i=1}^N\|\nu_i\|^2dt \nonumber \\
&&\le \frac{2\bar\sigma}{\sigma}\int_0^\infty \sum_{i=1}^N
\delta_i'Q_i\delta_i dt  
+ 2 \int_0^\infty \sum_{i=1}^N \|\nu_i\|^2dt <\infty. \qquad 
\label{L2convergence}
\end{eqnarray}

Next we prove statement (ii). First consider the disturbance and attack
free system comprised of the plant (\ref{eq:plant}) and the network of
observers (\ref{UP7.C.d}), when
$\xi\equiv 0$, $\xi_i\equiv 0$ and $f_i\equiv 0$  $\forall i$. In this
case, we also have $\hat f_i\equiv 0$,  
and $\nu_i\equiv 0$ since the system in Fig.~\ref{tracker} 
has zero initial conditions; see (\ref{Om.sys.general}). Asymptotic
stability of the system (\ref{e.2}) implies that in the disturbance and
attack free case, $z_i\to 0$, $\delta_i\to 0$
asymptotically. The latter property implies that $\|\hat
\epsilon_i-\hat f_i\|\to 0$, and since $f_i=\hat f_i\equiv 0$, then $\|\hat
\epsilon_i-f_i\|\to 0$ $\forall i$.   

When a disturbance or an attack input is present, i.e., if $\xi\not\equiv
0$ or, for at least one
$j$, $\xi_j\not \equiv 0$ or $f_j\not \equiv 0$, then it follows from
(\ref{objective.i.1}) that   
$\delta_i$, $z_i$, are $L_2$-integrable for all $i=1,\ldots,N$. Furthermore,
according to (\ref{ext.error}), $\dot \delta_i$ and $\dot z_i$ are also
$L_2$-integrable; this fact implies that $z_i\to 0$, $\delta_i\to 0$ and
$\varpi_i\to 0$ as $t\to 0$ for all $i=1,\ldots,N$. Then, to establish that
$\lim_{t\to\infty}\|\hat \varepsilon_i-f_i\|= 0$ $\forall i$, we consider
two cases. 

\emph{Case 1: } For all nodes $i$ which are not under attack, $f_i\equiv
0$ and $\hat f_i\equiv 0$. In this case, $\delta_i\to 0$ implies
$\hat\varepsilon_i=-\varpi_i\to 0$ asymptotically, and since $f_i\equiv 0$,
we have $\lim_{t\to\infty}\|\hat \varepsilon_i-f_i\|= 0$. 

\emph{Case 2: } 
When node $i$ is under attack, then $f_i\not\equiv 0$. At that node, we have 
$
\|\hat\varepsilon_i-f_i\|\le \|\varpi_i\|+\|\nu_i\|. 
$
Equation (\ref{eq:2}) states that
$\lim_{t\to\infty}\|\nu_i\|= 0$, and we have established previously that
$\varpi_j\to 0$ asymptotically for all $j=1,\ldots,N$, including $j=i$.
This implies that $\hat\varepsilon_i$ tracks $f_i$.  
asymptotically.
\hfill$\Box$

\begin{remark}
Part (i) of Theorem~\ref{L2.tracking} guarantees that each residual
output of the detector converges to the corresponding admissible attack
input $f_i$ in an $L_2$ sense. In part (ii), by taking into account the
properties of admissible biasing attack inputs of class $\mathcal{F}$, 
(of which the biasing 
attack inputs considered in~\cite{TSSJ-2015} are a special case), a sharper
asymptotic tracking behaviour of the residual variables $\hat\varepsilon_i$ is
obtained. This however requires a version of the condition
(\ref{objective.i.1}) to hold in which $\bar Q_i>0$ $\forall i$. In the sequel,
conditions will be given which guarantee this.
\hfill$\Box$
\end{remark}

We explain in the next section how the coefficients $\tilde L_i$,
$\tilde K_i$, $\check L_i$, and $\check K_i$ can be found to guarantee
satisfaction of the conditions stated in Problem~\ref{Prob1}. This will provide
a complete solution to the problem of detecting biasing attacks on 
distributed state observer networks under consideration. Further in
Section~\ref{sect.resilience}, it will be shown that the proposed detector
can also be used to negate effects of biasing attacks.   

\section{A vector dissipativity-based design of the attack detector}
\label{sec:lmi_design}

In the previous section we have recast the problem of attack detection
under consideration as a
problem of distributed stabilization of the large-scale system comprised of
subsystems~(\ref{ext.error}) via output
injection. References~\cite{HCN-2004,U6,LaU1} 
developed a vector dissipativity approach to solve this class of
problems. This approach will be applied here to obtain an algorithm for
constructing a state observer network to detect biasing attacks on
distributed filters. The idea behind this approach is to determine
the coefficients $\tilde L_i$, $\tilde K_i$, $\check L_i$, and $\check K_i$
for the error dynamics system (\ref{ext.error}) to ensure that each
subsystem (\ref{ext.error}) satisfies certain dissipation inequalities
\begin{equation}
\dot V_i+ 2\alpha_iV_i+ \delta_i'\check Q_i\delta_i+z_i'\tilde Q_iz_i \le \sum_{j\in
  \mathbf{V}_j}\pi_j V_j +\gamma^2\|\mathbf{w}_i\|^2,
\label{vec.Lyap}
\end{equation}
where $V_i(z_i,\delta_i)$ is a candidate storage function for
the error dynamics system (\ref{ext.error}), $\check Q_i\ge 0$, $\tilde
Q_i\ge 0$ are symmetric positive semidefinite matrices, and $\alpha_i>0$
and $\pi_i> 0$, $i=1, \ldots, N$, are constants 
selected so that $q_i\pi_i<2\alpha_i$. 

Unlike standard dissipation inequalities, the vector dissipation 
inequalities (\ref{vec.Lyap}) are coupled. Next, we show how
they can be used to establish input tracking properties of 
the distributed attack detector (\ref{ext.obs.nu.1.om}). It utilizes a
collection of quadratic storage functions
$V_i(z_i,\delta_i)=[z_i'~\delta_i']\mathbf{X}_i [z_i'~\delta_i']'$, 
with  $\mathbf{X}_i=\mathbf{X}_i'>0$.
 
\begin{lemma}\label{vec.dissip.lemma}
Suppose a set of matrices $\mathbf{X}_i=\mathbf{X}_i'>0$ and constants
$\alpha_i>0$, $\pi_i\in (0,2\alpha_i/q_i)$ can be
found which verify
the inequalities (\ref{vec.Lyap}). Then the collection of detectors
(\ref{ext.obs.nu.1.om}) has properties stated in 
Problem~\ref{Prob1}, with the following matrices $P$ and $Q_i$, $\bar Q_i$ in
(\ref{objective.i.1}):
\[
P=\gamma^{-2}\sum_{i=1}^N\mathbf{X}_i^{11},
\]
where  $\mathbf{X}_i^{11}$ is the upper left block in the partition of
$\mathbf{X}_i$ 
compatible with the dimensions of $z_i$ and $\delta_i$; and 
\begin{eqnarray}
&&Q_i=\check Q_i+\rho \lambda_{\mathrm{min}}(\mathbf{X}_i)I>0, \quad
\mbox{and} \nonumber \\ 
&&\bar
Q_i=\tilde Q_i+\rho \lambda_{\mathrm{min}}(\mathbf{X}_i)I>0,
\label{Qrho} 
\end{eqnarray}
where $\rho=\min_i (2\alpha_i-q_i\pi_i)>0$. 
\end{lemma}

The proof of the lemma is similar to the proof of the corresponding vector
dissipativity results in~\cite{U6,LaU1,U8}. For completeness, it is
included in the Appendix. 

\begin{remark}One appropriate candidate for $\pi_i$ is $\pi_i=
\frac{2\alpha_i}{q_i+1}$, where $q_i$ is the out-degree of the graph node
$i$. Clearly $\pi_i= \frac{2\alpha_i}{q_i+1}< \frac{2\alpha_i}{q_i}$, which
makes the value of $\pi_i$ a suitable candidate to be used in condition
(\ref{vec.Lyap}). \hfill$\Box$
\end{remark}   

We now present a method to  compute the coefficients $\tilde L_i$, $\tilde
K_i$, $\check L_i$, $\check K_i$ to satisfy 
the
dissipation inequalities~(\ref{vec.Lyap}). Let
\begin{eqnarray}
  \label{notation}
&&  
\mathbf{A}_i=\left[\begin{array}{cc} A & -F_i\Upsilon_i\\ 0 & \Omega_i 
  \end{array}\right], \quad 
\mathbf{B}_{1i}=\left[\begin{array}{c}F_i \\ \Gamma_i \end{array}\right], \quad 
\mathbf{B}_2= \left[\begin{array}{cc} -B & 0\\ 0 & 0 
  \end{array}\right], 
\nonumber \\
&&
\mathbf{D}_i= \left[\begin{array}{cc} D_i & \bar
    D_i \end{array}\right],\quad 
\mathbf{C}_i= \left[\begin{array}{ccc} C_i & 0 
  \end{array}\right], \quad
\mathbf{H}= \left[\begin{array}{ccc} H & 0 
  \end{array}\right], \quad \nonumber \\
&&
\mathbf{L}_i= \left[\begin{array}{c}\tilde L_i \\ \check L_i \end{array}\right], \quad
\mathbf{K}_i= \left[\begin{array}{c}\tilde K_i \\ \check
    K_i\end{array}\right], \quad
  \label{Qmu}
  \mathbf{Q}_i=\left[\begin{array}{cc} \tilde Q_i & 0\\ 0 & \check Q_i
    \end{array} \right].
\end{eqnarray}
Suppose $D_i$ and $\bar{D}_i$ satisfy the condition
\begin{equation}\label{E2i}
   \mathbf{E}_i\triangleq \mathbf{D}_i\mathbf{D}_i' = D_iD_i'+{\bar
     D_i}{\bar D_i}'> 0; 
\end{equation}
this is a standard  assumption made in 
$H_\infty$ control problems \cite{bacsar2008h}.

\begin{lemma}\label{LMI.lemma}
Suppose the digraph $\mathbf G$, the matrices
$\check Q_i=\check Q_i'\ge 0$, $\tilde Q_i=\tilde Q_i'\ge 0$ 
and the constants $\alpha_i>0$, $\pi_i\in(0,2\alpha_i/q_i)$, $i=1,\ldots,N$,
are such that the coupled linear matrix inequalities in \eqref{LMI} (on the
next page) with
respect to the variables $\mathbf{X}_i = \mathbf{X}_i'>0$ and $\mathbf{M}_i$, $i=1,\ldots,N$, are
feasible. Then choosing   
\begin{eqnarray}
\mathbf{K}_i &=& -\mathbf{X}_i^{-1} \mathbf{M}_i, \nonumber \\ 
\mathbf{L}_i&=&(\gamma^2\mathbf{X}_i^{-1}\mathbf{C}_i'-\mathbf{B}_2\mathbf{D}_i')
\mathbf{E}_i^{-1}
\label{L}
\end{eqnarray}
ensures that the condition \eqref{vec.Lyap} holds. 

\begin{figure*}[!t]
\centering
\begin{eqnarray}
\left[\begin{array}{cccccc}
\mathbf{S}_i & \mathbf{X}_i\mathbf{B}_{1i} &
\mathbf{X}_i\mathbf{B}_2(I-\mathbf{D}_i'\mathbf{E}_i^{-1}\mathbf{D}_i) & -\mathbf{M}_i\mathbf{H} & \ldots &
-\mathbf{M}_i\mathbf{H} \\
\mathbf{B}_{1i}'\mathbf{X}_i & -\gamma^2 I & 0 & 0  & \ldots & 0 \\
(I-\mathbf{D}_i'\mathbf{E}_i^{-1}\mathbf{D}_i)\mathbf{B}_2'\mathbf{X}_i & 0 & -\gamma^2I & 0  &
\ldots & 0 \\
 -\mathbf{H}'\mathbf{M}_i' & 0 & 0 & -\pi_{j_1}\mathbf{X}_{j_1} & \ldots &
 0 \\
\vdots & \vdots & \vdots & \ddots & \vdots & \vdots  \\
 -\mathbf{H}'\mathbf{M}_i' & 0 & 0 & 0 & \ldots &
 -\pi_{j_{p_i}}\mathbf{X}_{j_{p_i}}
\end{array}\right]<0, \label{LMI} 
\end{eqnarray}
\begin{eqnarray}
\mathbf{S}_i&=&
\mathbf{X}_i\left(\mathbf{A}_i+\alpha_i I +\mathbf{B}_2\mathbf{D}_i' \mathbf{E}_i^{-1} \mathbf{C}_i\right) + \left(\mathbf{A}_i+\alpha_i I +\mathbf{B}_2\mathbf{D}_i' \mathbf{E}_i^{-1}
  \mathbf{C}_i\right)'\mathbf{X}_i \label{S_i} 
+p_i\mathbf{M}_i\mathbf{H}+p_i\mathbf{H}'\mathbf{M}_i'+\mathbf{Q}_i - \gamma^2 \mathbf{C}_i'\mathbf{E}_i^{-1}\mathbf{C}_i. \nonumber
\end{eqnarray} 
\hrulefill		
\end{figure*}
\end{lemma}

The proof of this lemma is given in the Appendix. Combined with
Theorem~\ref{L2.tracking}
, this lemma
provides a complete result on the design of biasing attack detectors for the
distributed observer~(\ref{UP7.C.d}). This result is now formally
stated. The first part of the following theorem is concerned with detecting
general admissible attacks targeting any of the observer nodes, while the
second part particularizes this result to biasing attacks of class
$\mathcal{F}$, including bias injection attacks considered in~\cite{TSSJ-2015}. 

\begin{theorem}\label{L2.tracking.final}\label{assympt.tracking.final}\label{theorem}
Suppose 
the coupled linear matrix
inequalities in \eqref{LMI} with respect to the variables $\mathbf{X}_i =
\mathbf{X}_i'>0$ and $\mathbf{M}_i$, $i=1,\ldots,N$, are feasible. Then,
partitioning the matrices in (\ref{L}) to obtain $\tilde L_i$, $\tilde
K_i$, $\check L_i$, $\check K_i$, and letting $\bar L_i=\tilde L_i-L_i$,
$\bar K_i=\tilde K_i-K_i$ guarantees that for all admissible attack inputs, 
$\int_0^\infty
\|\hat\varepsilon_i- f_i\|^2dt<\infty$ $\forall i$. 
Furthermore, $\lim_{t\to\infty}\|\hat \varepsilon_i-f_i\|= 0$ for all
attack inputs $f_i$ of class $\mathcal{F}$. In particular, this conclusion
holds for all biasing attack inputs of the form  `a constant plus an
exponentially decaying transient'. 
\end{theorem}

\emph{Proof: }
The theorem follows from
Lemmas~\ref{vec.dissip.lemma} and~\ref{LMI.lemma} and
Theorem~\ref{L2.tracking}. 
\hfill$\Box$ 

The claim of Theorem~\ref{L2.tracking.final} involves the collection of
LMIs~\eqref{LMI} coupled in 
the variables $\mathbf{X}_i$ and $\gamma$. When the attack detector network
is designed offline, these LMIs can be solved in a routine manner using the
existing software. Also, the LMI problem \eqref{LMI} can be formulated within an
optimization framework where one seeks to determine a suboptimal level of
disturbance attenuation in condition~(\ref{objective.i.1}). 
It has been
shown in~\cite{WuLiUA1a} that a similar optimization problem 
can be solved in a distributed manner, where each network node computes its
own gain coefficients by communicating with its nearest
neighbours over a balanced graph. The algorithm was based on the well known
distributed optimization 
methods~\cite{BT-1989,Boyd-2010}. \cite{WuLiUA1a} considered the problem of
suboptimal disturbance attenuation stated in~\cite{U6}. However~\cite{U6}
and this paper have a common feature in that the original problem is
reduced to the problem of stabilization of an uncertain large-scale system
by output injection, and in both cases, the LMIs reflect vector
dissipativity properties of the error dynamics. This leads us
to suggest that the approach of~\cite{WuLiUA1a} can potentially be a
candidate to consider should one need to synthesize a distributed attack
detector network of the form~(\ref{ext.obs.nu.1.om}) online.

\begin{remark}\label{attack_bound}
This paper derives the attack detector for the worst-case scenario where potentially \emph{all} nodes can be compromised at the same time. We do not 
impose an upper bound on the number of nodes that can be 
compromised, and our result guarantees the detector performance for this
worst case scenario. 
However, it is not unreasonable to query whether the
performance of the proposed attack detector can be improved if one knows that
certain nodes are safe. First, we note that at the safe node we have
$f_i(t)\equiv 0$ and $\nu_i(t)\equiv 0$. Therefore, reducing the number of
compromised nodes will immediately manifest itself in the reduced total
energy of the `attack tracking' errors
$\sum_{i=1}^N\int_0^\infty\|\nu_i\|^2dt$ which bounds the energy in the
detection error; see (\ref{L2convergence}). Furthermore, if we know that 
certain node $i$ is safe, this knowledge can be captured by letting
$F_i=0$. In the case, it follows from (\ref{e.ext.nu.om}) that
$\epsilon_i(t)\equiv 0$, and the error $e_i$ satisfies the same equation as
the error $e_i$  of the original unbiased observer (however, this does not
mean that the two errors are identical since $e_j$ in (\ref{e.ext.nu.om})
can be biased). Also, we can choose the transfer function $G_i(s)=0$ for this
node, which means we can let $\Gamma_i=0$. As a result, the matrix
$\mathbf{B}_{1i}$ becomes a zero matrix, and this simplified modeling will
result in the detector error system at this node subjected to `less
uncertainty'; see equation~(\ref{enon.fixed.1}) in the Appendix. We expect that this should have an effect on the convergence rate of the detector and its
robustness. 
\end{remark}

\begin{remark}\label{mitra}
  Unlike~\cite{MS-2016,MS-arxiv}, it is not necessary for the network graph
  in this paper to be dense. It is assumed in \cite{MS-2016,MS-arxiv} that the communication graph must be very dense to allow removing
of suspicious nodes while preserving the connectivity between the
agents. Such a restrictive assumption is not required in this paper.
We take advantage of the dynamic model~(\ref{Om.sys.general})
  of biasing attack inputs. As a result, our
attack detection algorithm is based on a model based estimation technique which 
generally does not limit the number of affected nodes.
This approach contrasts with the approach in~\cite{MS-2016,MS-arxiv}, which makes no
assumptions about the type of the biasing signal but limits the number
of compromised nodes that can be tolerated.
  
\end{remark}

\begin{remark}\label{network.detectability}
  A question arises as to how the network structure plays a role in
  satisfying the LMIs~(\ref{LMI}). Ultimately, as is common in the
  $H_\infty$ control theory, the feasibility of the LMIs
  (\ref{LMI}) is related to detectability properties of the system
  (\ref{ext.error}), and we will explain in 
  Section~\ref{sect.detectability} how the network topology influences
  detectability of the detector network. 
\end{remark}

\section{Resilient distributed estimation}\label{sect.resilience}

Based on the foregoing analysis, we now show that equipping the
network of estimators (\ref{UP7.C.d}) with the attack
detectors (\ref{ext.obs.nu.1.om}) allows to obtain state estimates of the
plant (\ref{eq:plant}) that are resilient to admissible biasing
attacks. More precisely, consider the network 
of estimators (\ref{UP7.C.d}) augmented with the attack detectors
(\ref{ext.obs.nu.1.om}) and introduce `corrected' estimates 
\begin{equation}
  \label{xhat.corr}
  \doublehat{x}_i=\hat x_i+\hat e_i;
\end{equation}
here $\hat x_i$ is the `biased' estimate produced by the observer
(\ref{UP7.C.d}), and $\hat e_i$ is the correction term representing an
estimate of the error $e_i$ produced by the attack detector
(\ref{ext.obs.nu.1.om}). Note that the correction term is added at every
network node, so that each node of the augmented observer-detector system
(\ref{UP7.C.d}), (\ref{ext.obs.nu.1.om}) produces two estimates of the
plant state, $\hat x_i$ and $\doublehat{x}_i$.

Clearly, $x-\doublehat{x}_i=e_i-\hat e_i=z_i$. That is, $z_i$ is the error
associated with the estimate (\ref{xhat.corr}). It follows from
Theorem~\ref{assympt.tracking}, that for biasing attacks of class
$\mathcal{F}$, solving Problem~\ref{Prob1} with $\bar Q>0$ ensures that
this error vanishes asymptotically. That is, unlike
estimates $\hat{x}_i$ delivered by (\ref{UP7.C.d}) which become biased when
$f_i\not\equiv 0$, the corrected estimates $\doublehat{x}_i$ maintain
fidelity under attack. Furthermore, equation (\ref{objective.i.1}) provides a
bound on performance of the distributed estimator comprised of the node
estimators (\ref{UP7.C.d}), the attack detectors (\ref{ext.obs.nu.1.om}) 
and the outputs (\ref{xhat.corr}). This discussion is now summarized as the
following theorem. 

\begin{theorem}
\label{T2}
Consider the observer network (\ref{UP7.C.d}) augmented with the distributed
networked attack detector (\ref{ext.obs.nu.1.om}) whose coefficients $\bar
L_i$, $\bar K_i$, $\check L_i$, $\check K_i$ are obtained from the LMIs
\eqref{LMI} using the procedure described in
Theorem~\ref{L2.tracking.final}. Then the following statements hold
\begin{enumerate}[(a)]
\item
In the absence of disturbances and attack,
$\doublehat{x}_i\to x$ exponentially for all $i=1,\ldots,N$;

\item
In the presence of perturbations and biasing attacks of class $\mathcal{F}$, 
$\lim_{t\to \infty} \|x-\doublehat{x}_i\|= 0$ $\forall i$. Furthermore,
the estimation error $z_i=x-\doublehat{x}_i$ satisfies
(\ref{objective.i.1}) with $Q_i>0$, $\bar Q_i>0$ defined in (\ref{Qrho}). 
That is, $\doublehat{x}_i$ provides a resilient estimate of the plant when the
observer network is subject to a biasing attack. In
particular, this conclusion holds for all biasing attack inputs of the form 
`a constant plus an exponentially decaying transient'.      
\end{enumerate}
\end{theorem}
 
\emph{Proof: }
The conditions of the theorem guarantee that (\ref{vec.Lyap}) holds for
every $i$; see Lemma~\ref{LMI.lemma}. Furthermore, as was shown in
Lemma~\ref{L2.tracking.final}, this implies that statements (i) and
(ii) of Problem~\ref{Prob1} hold, with $Q_i>0$, $\bar Q_i>0$ defined in
(\ref{Qrho}). Finally, 
we have observed in the proof of Theorem~\ref{assympt.tracking} that since
according to (\ref{objective.i.1}), $\delta_i$, $z_i$ are $L_2$-integrable
for all $i$, we have $\dot z_i\in L_2$, for all $L_2$-integrable $\xi$,
$\xi_i$ and admissible $f_i$. This implies that $z_i\to 0$.~$\Box$

Theorem~\ref{T2} shows that the proposed attack detection
network is capable of mitigating biasing attacks on distributed state
estimation networks. Condition (\ref{objective.i.1}) characterizes its
performance under attack. Of course, when the system is attack free,
performance of the augmented observer-detector filter
(\ref{UP7.C.d}), (\ref{ext.obs.nu.1.om}) may be inferior to performance of
the original unbiased distributed filter (\ref{UP7.C.d.unbiased}), and we
do not propose $\doublehat x_i$ as a replacement for $\hat x_i$ in the
attack free situation. On the other hand, when some of the network
nodes are misappropriated and are subject to biasing attacks, the signals
$\doublehat x_i$ produced by the augmented observer-detector system
(\ref{UP7.C.d}), (\ref{ext.obs.nu.1.om}) are unbiased. This
shows that augmenting the observer network (\ref{UP7.C.d}) with the
network of attack monitors (\ref{ext.obs.nu.1.om}) provides a guarantee of
resilience, ensuring that the distributed observer remains functional
during hostile operating conditions. 
Of course, we do not suggest using inferior estimates $\doublehat{x}_i$ in an attack free situation. When the network is not under attack, we have $f_i=0$ and the state observer~(\ref{UP7.C.d}) produces unbiased estimates $\hat x_i$ of the plant state $x$ which are identical to the estimates produced by the original observer (\ref{UP7.C.d.unbiased}). In this case, we do not observe a performance degradation. The attack detector will produce zero residuals in this case. However when the network is subjected to a biasing attack, the residuals will deviate from zero. This will signal the presence of an attack. A threshold-based policy can then be devised to switch the observer outputs from the original estimates $\hat x_i$ to the resilient estimates $\doublehat{x}_i$. The design of such threshold-based policy is well studied in the fault detection literature, and we refer the reader to that literature; see e.g. \cite{Hwang2010}.

\section{Detectability of biasing attacks and relation to the network
  topology}\label{sect.detectability}

The role of the network topology in facilitating distributed estimation is
an interesting question which is under active investigation. For
networks of observers of the form (\ref{UP7.C.d.unbiased}),
conditions for detectability were obtained
in~\cite{U7b-journal,U7c}. In particular, this necessarily requires the pair
$(\bar{A}, [\bar{C}', \bar{H}']')$ to be detectable; here
$\bar{A}=I_N\otimes A$, $\bar{C}=\mbox{diag}[C_1,\cdots,C_N]$,
$\bar{H}=\mathcal{L}\otimes H$, and $\mathcal{L}$ is the $N\times N$
Laplacian matrix of the graph. It was shown in~\cite{U7b-journal,U7c}
that several factors affect the detectability of the network: 
(a) the decomposition of the network into components spanned by
trees, (b) 
the detectability properties of the pairs $(A,C_i)$, (c) the
observability properties of the pair $(A,H)$. From the results
in~\cite{U7b-journal,U7c}, for $(\bar{A},[\bar{C}', \bar{H}']')$ to be
detectable, each node must be able to
reconstruct from its interconnections with the neighbours the 
portion of the state information which cannot be obtained from its
local measurements. This makes estimation task feasible even when the
Laplacian matrix $\mathcal{L}$ has more than one zero eigenvalue. 
A general condition on the graph structure is that there should be a path
in the network from the sensors that can measure a certain portion of the states
to those that cannot measure this portion. So in general, if there are
several sensors that can measure the same portion of the state, it is not
necessary for them to be connected, and they provide this information to
other nodes in the subgraphs they belong to. More
recently, similar conclusions have been made in~\cite{PM-2017,MS-2018,WM-2017},
where somewhat more general data fusion schemes were considered using observers
whose dimension is greater than the dimension of the plant's state vector
$x$. 

In this section we build on the results in~\cite{U7b-journal,U7c} and
provide some insight into some fundamental attack input detectability
properties of the proposed distributed attack detector. 
    
Define $\bar{e}_i=[z_i'~\delta_i']'$ and let $\bar{e}$ be the vector of all
detector errors stacked together, $\bar e=[z,\delta]$, $z=[z_1', \ldots
z_N']'$, $\delta=[\delta_1',\ldots,\delta_N]'$. 
Then the disturbance and
attack-free detection error dynamics in~(\ref{e.2}) can be written in a
compact form,  
\begin{align}
\dot{\bar{e}} 
&= \left[\begin{array}{cc} \bar{A} & -\bar{F} \\  0 & \bar{\Omega}  \end{array}\right]\bar{e}
 + \left[\begin{array}{cc} -\tilde{L} & -\tilde{K} \\  -\check{L} & -\check{K} \end{array}\right] 
 \left[\begin{array}{cc} \bar{C} & 0 \\ \bar{H} &0 \end{array}\right]\bar{e},\label{e.2b}
\end{align} 
where $\tilde{L}=\mbox{diag}\{\tilde{L}_i\}$,
$\tilde{K}=\mbox{diag}\{\tilde{K}_i\}$,
$\bar{F}=\mbox{diag}\{F_i\Upsilon_i\}$,
$\check{L}=\mbox{diag}\{\check{L}_i\}$,
$\check{K}=\mbox{diag}\{\check{K}_i\}$ and
$\bar{\Omega}=\mbox{diag}\{\Omega_i\}$. Also, define 
\begin{align}
\mathcal{A} &= \left[\begin{array}{cc} \bar{A} & -\bar{F} \\  0 & \bar{\Omega}  \end{array}\right], \qquad \mathcal{C} = \left[\begin{array}{cc} \bar{C} & 0 \\ \bar{H} &0 \end{array}\right]. 
\end{align}
We conclude that for the system (\ref{ext.error}) to be stabilizable via
output injection, the pair $(\mathcal{A},\mathcal{C})$ must necessarily be
detectable. We now relate this condition to the detectability of the
pair $(\bar{A},[\bar{C}', \bar{H}']')$ of the original observer network
(\ref{UP7.C.d}).     

Let $s^*$ and $\bar\Delta(s^*)$ be an unstable eigenvalue and the corresponding
eigenspace of $\bar\Omega$. Define the following sets, 
\begin{eqnarray*}
\mathcal{D}(s^*)&=&\left\{y: y=\bar F\delta,~\delta\in
  \bar\Delta(s^*)\right\}, \\
\mathcal{Y}(s^*)&=&\left\{y: y=(\bar
  A-s^*I)z,~z\in \mathrm{Ker}\bar{C} \cap \mathrm{Ker}\bar{H}\right\}.  
\end{eqnarray*}

\begin{theorem}\label{detectability.theorem}
The pair $(\mathcal{A},\mathcal{C})$ is detectable if and only if the 
following conditions hold:
 \begin{enumerate}[(i)]
  \item 
  the pair $\left(\bar{A}, \left[\begin{array}{c} \bar{C} \\
        \bar{H} \end{array}\right]\right)$ is detectable;
\item
the pair $(\bar\Omega,\bar F)$ is detectable; and 
\item
For every unstable eigenvalue $s^*$ of $\bar\Omega$,
\begin{equation}
  \label{eq:condition}
 \mathcal{Y}(s^*)\cap
  \mathcal{D}(s^*)=\{0\}. 
\end{equation}
\end{enumerate}
\end{theorem}

\emph{Proof: }
The pair $(\mathcal{A},\mathcal{C})$ is detectable if and only if
\cite{Hespanha2009} 
\begin{align}
\mathrm{rank}\left[\begin{array}{c} \mathcal{A}-sI \\ \mathcal{C} \end{array} \right] = n, \qquad \forall s\in\mathbb C: \mathrm{Re}(s)\geq0.
\end{align}
In other words, $(\mathcal{A},\mathcal{C})$ is detectable if and only if
$\forall s\in\mathbb{C}: \mathrm{Re}(s)\geq0$, the following equations hold
only for $[z'~\delta']'=0$: 
\begin{align}\label{eq:detect1}
(\mathcal{A}-sI) \left[\begin{array}{c} z \\ \delta \end{array}\right]=0, \qquad \mathcal{C}\left[\begin{array}{c} z \\ \delta \end{array}\right]=0.
\end{align}
Expanding \eqref{eq:detect1} we obtain 
\begin{subequations}
\label{eq:fast}
\begin{align}
  &(\bar{A}-sI)z-\bar{F}\delta = 0,\label{eq:d1}\\
  &(\Omega-sI)\delta = 0,\label{eq:d2}\\
  &\bar{C}z = 0,\label{eq:d3}\\
  &\bar{H}z = 0.\label{eq:d4}
\end{align}
\end{subequations}

\emph{Sufficiency. } We now verify that under the conditions (i)--(iii) of
the theorem, (\ref{eq:fast}) hold only if $z=0$, $\delta=0$.  

First consider the case where $s$, $\mathrm{Re}(s)\ge 0$, is not an
eigenvalue of $\bar \Omega$. In this case, (\ref{eq:d2}) implies
$\delta=0$ and the remaining conditions (\ref{eq:fast}) read that 
\begin{subequations}
\label{eq:fast.1}
\begin{align}
  &(\bar{A}-sI)z = 0,\label{eq:d1.1}\\
  &\bar{C}z = 0,\label{eq:d3.1}\\
  &\bar{H}z = 0.\label{eq:d4.1}
\end{align}
\end{subequations}
It then follows from (i) and (\ref{eq:fast.1}) that $z=0$. Hence, if $s$,
$\mathrm{Re}(s)\ge 0$, is not an eigenvalue of $\bar \Omega$
then (\ref{eq:fast}) implies $z=0$, $\delta=0$.

Next, suppose $s=s^*$, where $s^*$ is an unstable eigenvalue of
$\bar\Omega$. In this case, \eqref{eq:d2} allows for both a zero and a nonzero
solution $\delta^*$. The case where $\delta^*=0$ has been considered
previously, it has led to the conclusion that $z=0$, $\delta=0$ is the only
solution to the system (\ref{eq:fast}). In the case where $\delta^*\neq 0$, we
conclude that $\delta^*$ is an eigenvector corresponding to $s^*$
and $\delta^*\in \bar\Delta(s^*)$. Furthermore, since $(\bar\Omega,\bar F)$
is detectable according to (ii), then $\bar F\delta^*\neq 0$. It then follows
from \eqref{eq:condition} that for any $z$ which satisfies \eqref{eq:d3}
and \eqref{eq:d4}, $(\bar A-s^*I)z-F\delta^*\neq 0$. Hence, (\ref{eq:fast})
cannot have a nonzero solution in this case as well. 

In summary, we conclude that the pair $(\mathcal{A},\mathcal{C})$
is detectable. 

\emph{Necessity. }
In this part of the proof $(\mathcal{A},\mathcal{C})$ is assumed to be
detectable. We now show that a violation of any of the conditions in
(i)--(iii) results in (\ref{eq:fast}) having a nonzero solution $(z,\delta)$.

Suppose that   $\left(\bar{A}, \left[\begin{array}{c} \bar{C} \\
        \bar{H} \end{array}\right]\right)$ is not detectable. Then there
  exists $z^*\neq 0$ which satisfies (\ref{eq:fast.1}). Substituting
  $z=z^*$   into (\ref{eq:fast}) results in the equations 
  \begin{equation}
    \label{eq:5}
    \bar F\delta=0, \quad  (\bar \Omega - s I)\delta =0,
  \end{equation}
which are satisfied with $\delta=0$. Thus, when $\left(\bar{A}, \left[\begin{array}{c} \bar{C} \\
        \bar{H} \end{array}\right]\right)$ is not detectable, then
  (\ref{eq:fast}) admits a nonzero solution $(z^*,0)$. This contradicts the
  assumption that $(\mathcal{A},\mathcal{C})$ is detectable.

Next, suppose $(\bar\Omega,\bar F)$ is not detectable. Let $s^*$ be an
unstable unobservable mode of $(\bar\Omega,\bar F)$, and let $\delta^*$ be 
a nonzero solution of (\ref{eq:5}) with $s=s^*$. Substituting 
$\delta=\delta^*$ and $z=0$ into (\ref{eq:fast}) shows that $(0,\delta^*)$
is a solution to (\ref{eq:fast}) when $s=s^*$. We have arrived at a 
contradiction with the assumption that $(\mathcal{A},\mathcal{C})$ is
detectable. 
 
Finally, suppose that $\bar\Omega$ has an unstable eigenvalue $s^*$ for
which the set $\mathcal{Y}(s^*)\cap
  \mathcal{D}(s^*)$ contains $y^*\neq 0$. This implies the existence of a
  nonzero $z^*\in\mathrm{Ker}\bar{C} \cap \mathrm{Ker}\bar{H}$ and a
  nonzero $\delta^*\in\Delta(s^*)$ such that
$
(\bar   A-s^*I)z^*=\bar F\delta^*=y^*.
$
Hence, we conclude that $(z^*,\delta^*)\neq 0$ 
satisfies (\ref{eq:fast}). Again, this conclusion is in contradiction with 
the assumption that $(\mathcal{A},\mathcal{C})$ is detectable.
\hfill $\Box$
 
Condition (i) can be related to detectability properties of each node and
properties of the network topology. As mentioned, 
 detectability of the pair $\left(\bar{A},  
    \left[\begin{array}{c} \bar{C} \\ \bar{H} \end{array}\right]\right)$
is related to the properties of the network graph, detectability properties
of the pairs $(A,C_i)$ and observability properties of $(A,H)$. We refer
the reader to~\cite{U7b-journal} for the analysis of this relationship. As far as
our analysis in this paper is concerned, we consider attacks on a
\emph{given} network of observers (\ref{UP7.C.d.unbiased}), therefore it is
reasonable to assume that the pair $\left(\bar{A},  
    \left[\begin{array}{c} \bar{C} \\ \bar{H} \end{array}\right]\right)$ is
  detectable, otherwise such a network will not be functional. 
The detectability of the pair $(\bar\Omega,\bar F)$ (condition (ii)) is
  immediately related to the detectability of every pair
  $(\Omega_i,F_i\Upsilon_i)$ $i=1,\ldots, N$. If the
attack tracking model does not guarantee this, this creates a possibility
for the attack detector to have an undetectable subspace. Estimation errors
within those subspaces cannot be monitored accurately, and the attacker can
exploit this fact to create  biased estimates.   
Condition (iii) is more subtle; it is directly related to the network
structure since it involves a set which is a linear transformation of a
subset of $\mathrm{Ker}\mathcal{L}\otimes 
H$; the latter set depends on the graph Laplacian  $\mathcal{L}$.
The failure to satisfy this condition will also lead to biased 
detector errors.

\section{Simulations}
\label{sec:simulations}
To illustrate the performance of the fault detection algorithm proposed in this
paper, we revisit the example in \cite{U6} where 
\begin{equation}
\begin{split}
&A = \left[\begin{array}{cccccc}
  0.3775 &      0 &      0 &         0 &      0 &       0 \\
  0.2959 & 0.3510 &      0 &         0 &      0 &       0 \\
  1.4751 & 0.6232 & 1.0078 &         0 &      0 &       0 \\
  0.2340 &     0  &      0 &    0.5596 &      0 &       0 \\
       0 &      0 &      0 &    0.4437 & 1.1878 & -0.0215 \\
       0 &      0 &      0 &         0 & 2.2023 &  1.0039 \\
\end{array}\right], \\
& B = 0.1I_{6\times6}, \quad H=I_{6\times6},\\
& D_i = 0_{2\times6}, \quad \bar{D}_i = 0.01 I_2, \quad \forall i=1,\ldots,6,
\end{split}
\end{equation}
and each sensor $i$ measures the $i$-th and $(i+1)$-th coordinates of the
state vector, with sensor 6 measuring the 6th and 1st
coordinates. Therefore, for example for the 4th sensor, we have 
\begin{equation}
C_{4} = \left[\begin{array}{cccccc}
  0 & 0 & 0 & 1 &  0 & 0 \\
  0 & 0 & 0 & 0 &  1 & 0 \\
\end{array}\right].
\end{equation}
The reason why this example is chosen is that all the pairs $(A,
C_i),~i=1,\cdots,6$ in this example are not observable, and $A$ is
anti-stable which ensures that at every node of the network, the
unobservable modes of $A$ are not detectable. In \cite{U6}, a distributed
observer was constructed for this system which consisted of $N=6$ observer
nodes interconnected over a simple circular digraph, with
$\mathbf{V}=\{1,2,3,4,5,6\}$ and
$\mathbf{E}=\{(6,1),(1,2),(2,3),(3,4),(4,5),(5,6)\}$.
 
We assume that scalar biasing attacks can be applied at any
node of the observer network, and assume $F_i=[1~1~1~1~1~1]'$ $\forall i=1,
\ldots,6$. We limit attention to the special case of bias inputs
admissible with $G_i(s)=\frac{1}{s+2\beta_i}I_6$, $\beta_i>0$ and 
design an $L_2$-tracking detector based on Theorem~\ref{theorem}. We let 
$\beta_i$, $i=1,\cdots, 6$ in \eqref{omega} be $\beta_i = 10$. Letting all
$\alpha_i=2$ and $\gamma^2=0.5$, it was found using the YALMIP software
\cite{Lofberg2004} that the LMI problem in \eqref{LMI} was feasible. The
LMI variables $\mathbf{K}_i$ and $\mathbf{L}_i$ in \eqref{L} were
calculated using YALMIP, and then using \eqref{notation}, the values for
$\tilde{L}_i$, $\tilde{K}_i$, $\check L_i$ and $\check K_i$ were
obtained. Using the gain values $L_i$ and $K_i$ of the observer in
\eqref{UP7.C.d.unbiased} obtained in the example in \cite{U6}, we then
calculated the observer gain values in \eqref{ext.obs.nu.1.om}, $\bar
L_i=\tilde{L}_i-L_i$ and $\bar K_i=\tilde{K}_i-K_i$.

To illustrate performance of the obtained attack detectors
(\ref{ext.obs.nu.1.om}) and the
corresponding resilient estimators~(\ref{UP7.C.d}), (\ref{xhat.corr}), the
system was simulated using Matlab. The initial
conditions of the plant (\ref{eq:plant}) were chosen randomly, and the
process and measurement disturbances were selected to be broadband white
noises of intensity~1. An attack signal $f_2$ was applied at node 2 at time
$t=2s$ which lasted for $5s$. During this time, the value of $f_2(t)$ in
\eqref{UP7.C.d} changes from zero at $t=2s$ to the value of 5 and becomes
zero again at $t=7s$. 

\begin{figure} [h]
  \includegraphics[width=0.4\textwidth]{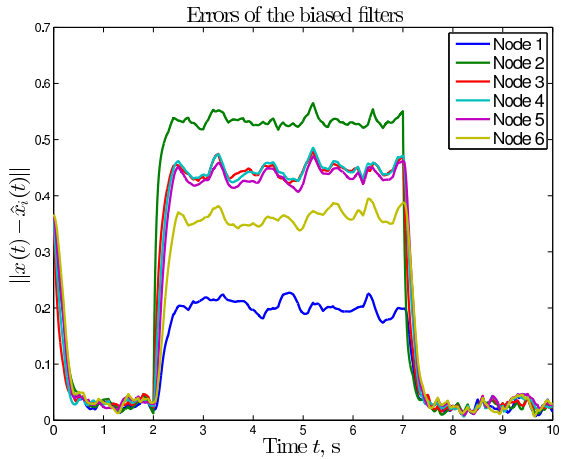}
  \caption{Norms of the errors of the observers (\ref{UP7.C.d}).}
  \label{fig:f1}
\end{figure}

\begin{figure} [h]
  \includegraphics[width=0.4\textwidth]{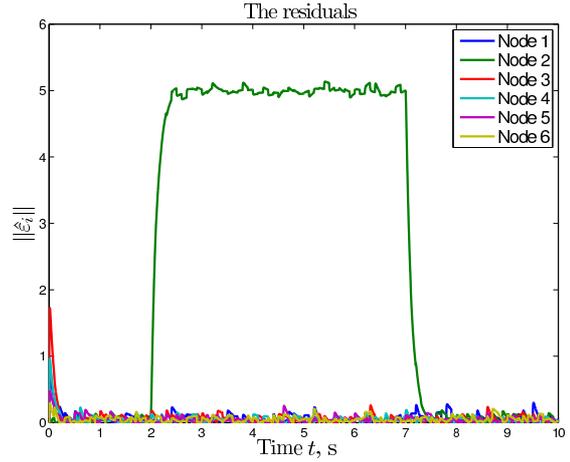}
  \caption{Norms of the residual outputs $\hat\varepsilon_i(t)$ of the
    attack detectors (\ref{ext.obs.nu.1.om}). All residuals except that at the
    misappropriated node 2, have low amplitude and only respond to
    disturbances in the system. On the other hand, the residual
    $\hat\varepsilon_2(t)$ is able to detect and track the biasing input
    $f_2(t)$. }  
  \label{fig:f3}
\end{figure}

\begin{figure} [h]
  \includegraphics[width=0.4\textwidth]{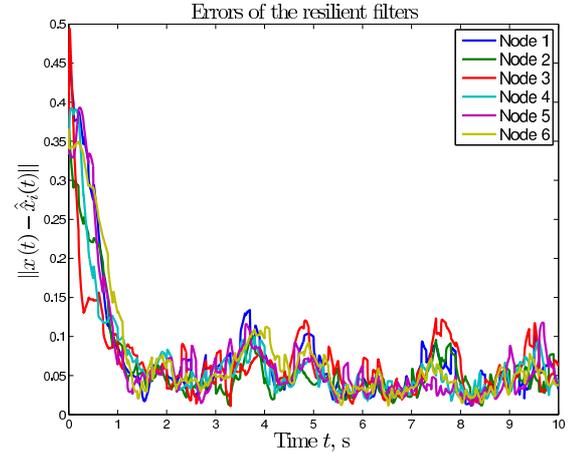}
  \caption{Norms of the errors of the resilient
    estimates~(\ref{xhat.corr}), produced using the biased observers
    (\ref{UP7.C.d}) augmented with the attack
    detectors~(\ref{ext.obs.nu.1.om}).}  
  \label{fig:f2}
\end{figure}

\begin{figure} [h]
  \includegraphics[width=0.4\textwidth]{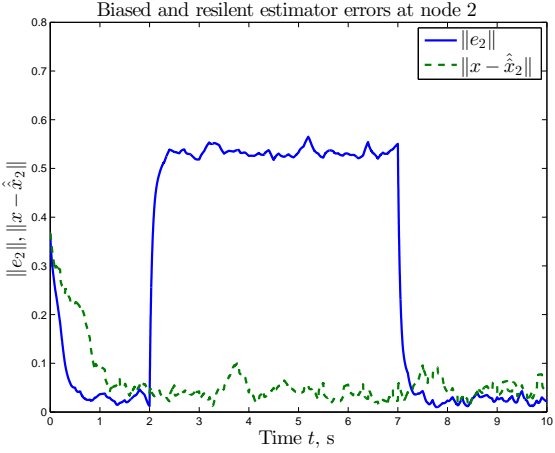}
  \caption{Norms of the errors of the biased and resilient estimates at
    the misappropriated node 2. The error of the resilient observer
    incorporating the attack detector is substantially lower than the error
    of the corresponding observer of the biased network. The estimate
    (\ref{xhat.corr}) at this node appears to be not affected by the
    attack, it maintains roughly the same level of accuracy during the
    attack as before and after the attack. However, the error is somewhat
    greater than the error of the original observer from~\cite{U6} when it
    operates normally.}
  \label{fig:f4}
\end{figure}

Figures~\ref{fig:f1}--\ref{fig:f4} show the errors exhibited by the obtained attack detectors (\ref{ext.obs.nu.1.om}) and the corresponding biased and resilient observers~(\ref{UP7.C.d}) and~(\ref{xhat.corr}), respectively, in response to this attack.  
 
\begin{figure} [h]
  \includegraphics[width=0.4\textwidth]{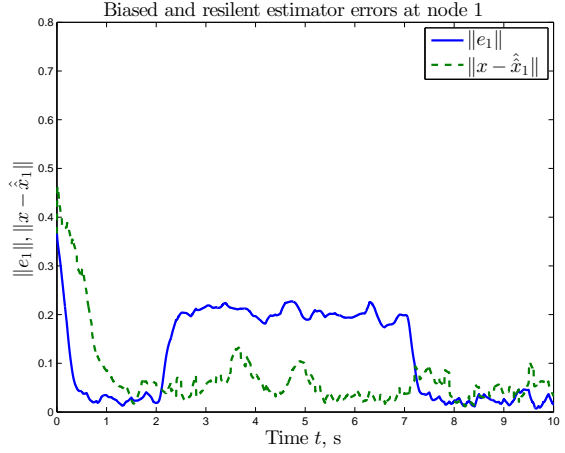}
  \caption{Norms of the errors of the biased and resilient estimates at
    node 1, which is most distant from the misappropriated node. Even at
    the most distant node, the error of the resilient observer incorporating
  the attack detector is almost half of the error of the corresponding
  observer of the biased network. The error appears to be unaffected by the
attack, although it is somewhat greater than the error of the original
observer from~\cite{U6} when it operates normally.}  
  \label{fig:f5}
\end{figure}

It can be seen in Fig.~\ref{fig:f1}, all nodes in the system are affected by the attack, and the estimation errors at every node become biased during the time interval $2\leq t\leq 7$. As expected, the biasing effect of the attack is most prominent at node 2, and node 1 is least affected, as $\|x(t)-\hat{x}_1(t)\|<\|x(t)-\hat{x}_i(t)\|< \|x(t)-\hat{x}_2(t)\|$ for $i=3,\cdots,6$ and for almost all $t\in[2,7]$.
However, Fig~\ref{fig:f3} shows that the attack detectors~(\ref{ext.obs.nu.1.om}) are able to reliably identify the source
of attack and track the attack input quite accurately. This figure shows
that $\hat \varepsilon_2(t)$ changes at $t=2 s$ and $t=7 s$ indicating an
attack at node 2, while other residual variables $\hat \varepsilon_i(t)$,
$i\neq 2$ appear to be unaffected by the attack. Also, the estimates
$\doublehat x_i$ computed according to (\ref{xhat.corr}) show much greater
resilience to the attack, compared with $\hat x_i$. Although $\doublehat x_i$
tend to be somewhat less accurate than $\hat x_i$ under normal conditions,
their error appear to be not affected by the attack; see
Fig.~\ref{fig:f2}. To further illustrate this point, Figures~\ref{fig:f4}
and \ref{fig:f5} compare the errors of the two observers at the most
affected node 2 and the least affected node~1. As 
one can see, in both cases the estimates $\doublehat x_i$ appear to be 
unaffected by the attack.
      
\section{Conclusion} \label{sec:conclusion}
The paper is concerned with the problem of distributed attack detection in
sensor networks. We consider a group of consensus-based distributed
estimators and assume that the estimator dynamics are under attack. Then we
propose a distributed attack detector which allows for an uncertainty in
the sensors and the plant model, as well as a range of bias attack inputs,
and show that the proposed attack detector can detect an biasing attack and
identify the misappropriated node. Also, we show that these detectors can
be used to compensate the biasing effect of the attack, once it is
detected. Although under normal circumstances, the proposed resilient
estimates are less accurate than the estimates produced by the original
network, they show superior resilience to the attack, in that they
asymptotically converge to the state of the plant under a broad range of
$L_2$ integrable perturbations and biasing attack inputs. The limitation of
the proposed scheme lies in the assumption that in principle, admissible
attack inputs can be tracked using a low-pass filter and that the tracking
error is $L_2$ integrable. This restricts the class of attack inputs that
can be detected and countered using our approach. Future effort will be
directed towards relaxing this assumption. 

Another future problem is to consider link failures under
  denial of service attacks which aim to disrupt the normal flow of information
  within the network. Sparse networks are more likely to fail under a
  jamming attack, and for the observer to maintain resilience, additional
  connectivity within the network may be required. This contrasts with the
  problem considered in this paper where the attacker relies on dense
  connectivity to spread the biased $\hat x_i$ across the network. 
  In this situation, sparse topologies appear to be beneficial for the
  defender. An 
  interesting problem would be to determine which strategy is more
  beneficial for the attacker facing a particular network (biasing, jamming
  or a combination of both), and which network structure provides for the
  best resilient performance under this strategy. We leave this challenging
  problem for future research.

\section*{Acknowledgement}
The authors would like to thank the Reviewers and the Associate Editor for their constructive comments.
The authors also thank G.~Seyboth for providing his
paper~\cite{SA-2015}. 

\section*{Appendix}
\subsection{Proof of equation (\ref{eq:2})}

Observe that an input $f_i$ of class $\mathcal{F}$ has a Laplace
transform of the form 
$
f_i(s)=\frac{R_0}{s}+\sum_k\frac{R_k}{(s+p_k)^{l_k}}
$
with  $\mathrm{Re}(p_k)<0$, $l_k\ge 1$, $\forall k$. By assumption,
$(I+\frac{1}{s}G_i(s))^{-1}$ has all its poles in the region
$\mathrm{Re}(s)<0$, therefore  $\forall f_i\in\mathcal{F}$,     
\begin{eqnarray*}
  \nu_i(s)=-(I+\frac{1}{s}G_i(s))^{-1}f_i(s)=
  \frac{\hat R_0}{s}+\sum_k\frac{\hat R_k}{(s+\hat p_k)^{\hat l_k}}  
\end{eqnarray*}
with  $\mathrm{Re}(\hat p_k)<0$, $\hat l_k\ge 1$, $\forall k$. This time
the summation is carried out 
over the joint set of poles which includes stable poles of both
$(I+\frac{1}{s}G_i(s))^{-1}$ and $f_i(s)$. Hence $\lim_{t\to
  \infty}\nu_i(t)$ exists. Furthermore,
\[
\|s\nu_i(s)\|\le \|(I+\frac{1}{s}G_i(s))^{-1}\|\cdot\|sf_i(s)\|
\]
and $\lim_{s\to 0}\|sf_i(s)\|=\lim_{t\to\infty}\|f_i(t)\|<\infty$. Then 
according to the final value theorem,
\begin{eqnarray*}
 \lim_{t\to\infty}\|f_i(t)-\hat f_i(t)\| 
&\le& \lim_{s\to 0}\left( \|(I+\frac{1}{s}G_i(s))^{-1}\|\cdot\|sf_i(s)\|\right) =0.
\end{eqnarray*}

\subsection{Proof of Lemma~\ref{vec.dissip.lemma}}
Let $V=\sum_{i=1}^NV_i$. Adding the inequalities (\ref{vec.Lyap}) and
selecting $\pi_i<\frac{2\alpha_i}{q_i}$ will result in
\begin{eqnarray}
\lefteqn{\dot{V}+ \sum_{i=1}^N(\delta_i'Q_i\delta_i+z_i'\tilde
  Q_iz_i)} && \nonumber \\ 
&& \le 
-\rho V+
\gamma^2\sum_{i=1}^N (\|\xi\|^2+\|\xi_i\|^2+\|\nu_i\|^2);
\label{Lyap.global}
\end{eqnarray}
here 
$\rho=\min_i(2\alpha_i-q_i\pi_i)>0$. This implies that when $\xi=0$ and $f_i=0$, $\xi_i=0$ $\forall i$, then
$
\dot V<-\rho V,
$
and since $\mathbf{X}_i>0$, we have $z_i\to 0$, $\delta_i\to 0$
exponentially. That is, condition (i) of Problem~\ref{Prob1} is
established.    

Also, when at least one of the signals $\xi$,  $\xi_i$  or $f_i$ is not equal
to zero (the latter is equivalent to $\nu_i\not\equiv 0$), then it follows
from (\ref{Lyap.global}) that with $Q_i$, $\bar Q_i$ defined in (\ref{Qrho}),
\[
\begin{split}
\sum_{i=1}^N &\int_0^T(\delta_i'Q_i\delta_i+z_i'\bar Q_iz_i)dt 
\le \sum_{i=1}^N \left[V_i(z_i(0),\delta_i(0)) \right.\\
&+ \left.\gamma^2
\int_0^T(\|\xi\|^2+\|\xi_i\|^2+\|\nu_i\|^2)dt\right].  
\end{split}
\]
Note that $V_i(z_i(0),\delta_i(0))=x_0'\mathbf{X}_i^{11}x_0$. Hence
(\ref{objective.i.1}) also holds with
$P=\gamma^{-2}\sum_{i=1}^N\mathbf{X}_i^{11}$ and $Q_i$, $\bar Q_i$ defined in (\ref{Qrho}). 

\subsection{Proof of Lemma~\ref{LMI.lemma}}
With the notation (\ref{notation}) and letting $\mu_i=[z_i'~\delta_i']'$,
the system (\ref{ext.error}) can be represented in the form
\begin{eqnarray}
   \dot{\mu}_i 
  &=& (\mathbf{A}_i - \mathbf{L}_i \mathbf{C}_i)\mu_i
  +\sum_{j\in\mathbf{V}_i}\mathbf{K}_i \mathbf{H}(\mu_j-\mu_i)
             \nonumber \\  
 &+&\mathbf{B}_{1i} \nu_i-(\mathbf{B}_2+\mathbf{L}_i \mathbf{D}_i)w_i(t),
  \label{enon.fixed.1} \qquad \\
  \mu_i(0)&=&\left[\begin{array}{c}z_i(0) \\ \delta_i(0)
    \end{array}\right], \quad 
w_i\triangleq \left[\begin{array}{c}\xi \\
    \xi_i \end{array}\right]. \nonumber 
\label{U8.vi.fixed} 
\end{eqnarray}
To establish the vector dissipativity properties of the system
(\ref{enon.fixed.1})  we proceed as in~\cite{U8,LaU1}. 

By pre-multiplying
and post-multiplying the matrix inequality \eqref{LMI} by $[\mu_i'~ \phi_1'~
\phi_2'~ \mu_{j_1}'~ \cdots \mu_{j_{p_i}}']$ and its transpose we obtain 
\begin{eqnarray}
&\hspace{-4ex}0>& 
2\mu_i'\mathbf{X}_i(\mathbf{A}_i+\alpha_i I -\mathbf{L}_i
\mathbf{C}_i)\mu_i +2\mu_i' \mathbf{X}_i\mathbf{K}_i \mathbf{H}\sum_{j\in \mathbf{V}_i} \mu_j 
\nonumber \\
&& +2\gamma^2\mu_i'\mathbf{C}_i'\mathbf{E}_i^{-1} \mathbf{C}_i \mu_i-2p_i\mu_i' \mathbf{X}_i\mathbf{K}_i \mathbf{H}\mu_i \nonumber \\
&&
+\mu_i'\mathbf{Q}_i\mu_i - \gamma^2\mu_i'\mathbf{C}_i'\mathbf{E}_i^{-1}\mathbf{C}_i\mu_i - \gamma^2\|\phi_1-\frac{1}{\gamma^2}\mathbf{B}_{1i}'\mathbf{X}_i\mu_i\|^2 \nonumber \\ 
&& + \frac{1}{\gamma^2}\mu_i'\mathbf{X}_i\mathbf{B}_{1i} \mathbf{B}_{1i}'\mathbf{X}_i\mu_i -\gamma^2\|\phi_2-\frac{1}{\gamma^2}\left(I-\mathbf{D}_i'\mathbf{E}_i^{-1}\mathbf{D}_i\right)\mathbf{B}_2'\mathbf{X}_i\mu_i\|^2  \nonumber \\
&& + \frac{1}{\gamma^2}\mu_i'\mathbf{X}_i\mathbf{B}_2\left(I-\mathbf{D}_i'\mathbf{E}_i^{-1}\mathbf{D}_i\right)\mathbf{B}_2'\mathbf{X}_i\mu_i
- \sum_{j\in\mathbf{V}_i}\pi_j\mu_j'\mathbf{X}_j\mu_j.\nonumber
\end{eqnarray} 
Note that $\left(I-\mathbf{D}_i'\mathbf{E}_i^{-1}\mathbf{D}_i\right)$ is a projection matrix and thus 
$\left(I-\mathbf{D}_i'\mathbf{E}_i^{-1}\mathbf{D}_i\right)\left(I-\mathbf{D}_i'\mathbf{E}_i^{-1}\mathbf{D}_i\right)=
\left(I-\mathbf{D}_i'\mathbf{E}_i^{-1}\mathbf{D}_i\right).
$ Furthermore, it can be shown by direct calculations that 
$\frac{1}{\gamma^2}\mathbf{B}_2\left(I-\mathbf{D}_i'\mathbf{E}_i^{-1}\mathbf{D}_i\right)\mathbf{B}_2' =\frac{1}{\gamma^2}(\mathbf{B}_2+\mathbf{L}_i \mathbf{D}_i)(\mathbf{B}_2+\mathbf{L}_i \mathbf{D}_i)'
-\gamma^2 \mathbf{X}_i^{-1}\mathbf{C}_i'\mathbf{E}_i^{-1}\mathbf{C}_i\mathbf{X}_i^{-1}.
$ 
Hence for any vector  $[\mu_i~ \mu_{j_1}~ \cdots \mu_{j_p}]\neq0$, letting
$\phi_2 =
\frac{1}{\gamma^2}\left(I-\mathbf{D}_i'\mathbf{E}_i^{-1}\mathbf{D}_i\right)\mathbf{B}_2'\mathbf{X}_i\mu_i
$ and $\phi_1=\frac{1}{\gamma^2}\mathbf{B}_{1i}'\mathbf{X}_i\mu_i$  leads to
the inequality 
\begin{eqnarray}
&0>&
2\mu_i'\mathbf{X}_i(\mathbf{A}_i+\alpha_i I -\mathbf{L}_i
\mathbf{C}_i)\mu_i + \mu_i'\mathbf{Q}_i\mu_i - \sum_{j\in\mathbf{V}_i} \pi_j\mu_j'\mathbf{X}_j\mu_j  \nonumber \\ 
&& +\gamma^2\|\nu_i-\frac{1}{\gamma^2}(\mathbf{B}_{1i})'\mathbf{X}_i\mu_i\|^2+2\mu_i'\mathbf{X}_i\mathbf{B}_{1i}\nu_i - \gamma^2\|\nu_i\|^2  \nonumber \\
&& +\gamma^2\|w_i+\frac{1}{\gamma^2}(\mathbf{B}_2+\mathbf{L}_i
\mathbf{D}_i)'\mathbf{X}_i\mu_i\|^2-\gamma^2\|w_i\|^2 
\nonumber \\
&& - 2\mu_i'\mathbf{X}_i(\mathbf{B}_2+\mathbf{L}_i \mathbf{D}_i)w_i +
2\mu_i'
\mathbf{X}_i\mathbf{K}_i \mathbf{H}\sum_{j\in \mathbf{V}_i}(\mu_j-\mu_i).  \nonumber
\end{eqnarray} 
Then it follows from the above inequality that 
\begin{eqnarray}
\hspace{-4ex}\dot V_i\le -2\alpha_i\mu_i'\mathbf{X}_i\mu_i - \mu_i'\mathbf{Q}_i\mu_i  + \gamma^2\|\nu_i\|^2 + \gamma^2\|w_i\|^2 + \sum_{j\in\mathbf{V}_i} \pi_j\mu_j'\mathbf{X}_j\mu_j.\nonumber 
\end{eqnarray}
Thus 
for all $i=1,\cdots,N$ the inequality \eqref{vec.Lyap} holds.

\newcommand{\noopsort}[1]{} \newcommand{\printfirst}[2]{#1}
  \newcommand{\singleletter}[1]{#1} \newcommand{\switchargs}[2]{#2#1}

\end{document}